\documentclass[preprint]{elsarticle}

\usepackage{amsmath}
\usepackage{xspace}

\newenvironment{cfigure}[1][tbp]{\begin{figure}[#1]\centering}{\end{figure}}
\newenvironment{cfigure1c}[1][tbp]{\begin{figure*}[#1]\centering}{\end{figure*}}

\newcommand{\fig}[1]{Fig.~\ref{#1}}
\newcommand{\tab}[1]{Table~\ref{#1}}

\newcommand{\secref}[1]{Sec.~\ref{#1}}
\newcommand{\eqnref}[1]{Eq.~\eqref{#1}}

\newcommand{\Fig}[1]{Figure~\ref{#1}}
\newcommand{\Tab}[1]{Table~\ref{#1}}

\newcommand{\figrange}[2]{Figs.~\ref{#1}--\ref{#2}}
\newcommand{\Figrange}[2]{Figures~\ref{#1}--\ref{#2}}
\newcommand{\eqnrange}[2]{Eqs.~\ref{#1}--\ref{#2}}

\newcommand{\En}{\ensuremath{E_{\text{n}}}\xspace}
\newcommand{\Ep}{\ensuremath{E_{\text{p}}}\xspace}

\newcommand{\Enprime}{\ensuremath{E_{\text{n}^{\prime}}}\xspace}
\newcommand{\Dxzero}{\ensuremath{\Delta x_{0}}\xspace}
\newcommand{\Dyzero}{\ensuremath{\Delta y_{0}}\xspace}
\newcommand{\Dzzero}{\ensuremath{\Delta z_{0}}\xspace}
\newcommand{\Dtzero}{\ensuremath{\Delta t_{0}}\xspace}
\newcommand{\Dnzero}{\ensuremath{\Delta n_{0}}\xspace}
\newcommand{\Dxone}{\ensuremath{\Delta x_{1}}\xspace}
\newcommand{\Dyone}{\ensuremath{\Delta y_{1}}\xspace}
\newcommand{\Dzone}{\ensuremath{\Delta z_{1}}\xspace}
\newcommand{\Dtone}{\ensuremath{\Delta t_{1}}\xspace}
\newcommand{\Dnone}{\ensuremath{\Delta n_{1}}\xspace}
\newcommand{\dsep}{\ensuremath{d_{10}}\xspace}
\newcommand{\tsep}{\ensuremath{t_{10}}\xspace}
\newcommand{\Ddsep}{\ensuremath{\Delta \dsep}\xspace}
\newcommand{\Dtsep}{\ensuremath{\Delta \tsep}\xspace}
\newcommand{\DEp}{\ensuremath{\Delta \Ep}\xspace}
\newcommand{\DEn}{\ensuremath{\Delta \En}\xspace}
\newcommand{\Dtheta}{\ensuremath{\Delta \theta}\xspace}
\newcommand{\Dalpha}{\ensuremath{\Delta \alpha}\xspace}

\newcommand{\genunit}[2]{\ensuremath{#1~\text{#2}}\xspace}
\newcommand{\meters}[1] {\genunit{#1}{m}}
\newcommand{\cm}[1]     {\genunit{#1}{cm}}
\newcommand{\kev}[1]    {\genunit{#1}{keV}}
\newcommand{\kevee}[1]    {\genunit{#1}{keVee}}
\newcommand{\mev}[1]    {\genunit{#1}{MeV}}
\newcommand{\mevnoarg}  {\ensuremath{\mathrm{MeV}}}
\newcommand{\mcub}[1]   {\ensuremath{#1~\mathrm{m}^{3}}\xspace}

\newcommand{\Cftft}{\ensuremath{^{252}\text{Cf}}\xspace}

\usepackage{textcomp}
\usepackage[mathlines]{lineno}
\usepackage{siunitx}
\usepackage{url}

\journal{Nuclear Instruments and Methods in Physics A}

\begin{document}

\begin{frontmatter}

\title{Single-Volume Neutron Scatter Camera for High-Efficiency Neutron Imaging and Spectroscopy}

\author[snl]{Joshua Braverman}
\author[snl]{James Brennan}
\author[snl]{Erik Brubaker\corref{corauth}}
\author[snl]{Belkis Cabrera-Palmer}
\author[snl,orst]{Steven Czyz}
\author[snl]{Peter Marleau}
\author[ncsu]{John Mattingly}
\author[snl]{Aaron Nowack}
\author[snl]{John Steele}
\author[snl]{Melinda Sweany}
\author[snl,ncsu]{Kyle Weinfurther}
\author[snl]{Eli Woods}

\address[snl]{Sandia National Laboratories, Livermore, CA}
\address[ncsu]{North Carolina State University, Raleigh, NC}
\address[orst]{Oregon State University, Corvallis, OR}

\cortext[corauth]{Corresponding author: ebrubak@sandia.gov}

\begin{abstract}
Neutron detection provides an effective method to detect, locate, and 
characterize sources of interest to nuclear security applications. Current 
neutron imaging systems based on double-scatter kinematic reconstruction 
provide good signal vs.\ background discrimination and spectral capability, 
but suffer from poor sensitivity due to geometrical constraints.
This weakness can be overcome if both 
neutron-proton scattering interactions are detected and resolved within one 
large contiguous active detector volume. We describe here a maximum 
likelihood approach to event reconstruction in a single-volume system with no 
optical segmentation and sensitivity to individual optical photons on the 
surfaces of the scintillator. We present results from a Geant4-based 
simulation establishing the feasibility of this single-volume neutron scatter 
camera concept given notional performance of existing photodetector and
readout technologies.
\end{abstract}


\end{frontmatter}

\section{Introduction}
\label{s:introduction}
Fission-energy neutrons are a sensitive and specific signature of special 
nuclear material, due to their low and stable natural backgrounds, their penetrating 
nature, and the scarcity of benign neutron-emitting materials.  As such, 
neutron imaging has the potential to be an important tool in a range of 
nuclear security applications, including arms control treaty verification, 
emergency response, cargo screening, and standoff search. In particular, 
neutron scatter cameras (NSC) have become an established 
technology for fission-energy neutron imaging and spectroscopy~\cite{Ryan_1993,Vanier_2006, Marleau_2007,goldsmith_2016,Poitrasson-Riviere_2014}.  Based on the kinematic 
reconstruction of neutron double scatters, NSC systems provide excellent
event-by-event directional information for signal-to-background discrimination, 
reasonable imaging resolution, and good energy resolution when compared to 
competing imaging technologies~\cite{Poitrasson-Riviere_2014,Mascarenhas_2009,Brennan_2011,Weinfurther_2017}. 

The operational principle of an NSC is based on the neutron elastically 
scattering on hydrogen at least twice in the detector's active material, typically an 
organic scintillator. The incoming kinetic energy \En is obtained from 
\begin{equation}
\label{eq:kin1}
\En=\Enprime+\Ep
\end{equation}
as the sum of the energy \Ep deposited in the first scatter, which is assumed 
to generate a recoiling proton, plus the scattered neutron energy \Enprime. 
This in turn is calculated from
\begin{equation}
\label{eq:kin2}
\Enprime=\frac{1}{2}m_{\text{n}}\left(\frac{\dsep}{\tsep}\right)^{2},
\end{equation}
using the measured distance \dsep and time \tsep between the first two 
scatters. These quantities also allow for a calculation of the scattering angle
$\theta$ defining a cone containing the incoming neutron direction:
\begin{equation}
\label{eq:kin3}
\theta=\arccos\left(\sqrt{\frac{\Enprime}{\En}}\right).
\end{equation}
Most existing NSC systems consist of arrays of spatially-separated organic 
scintillator cells in order to isolate in two distinct cells the neutron's 
two elastic scatters~\cite{Ryan_1993,Vanier_2006, Marleau_2007,goldsmith_2016,Poitrasson-Riviere_2014}.
This requirement, however, has poor geometrical efficiency, constituting the primary 
drawback of current NSC systems.

The design concept of a single-volume neutron 
scatter camera (SVSC), in which both neutron scatters occur in the same large 
active volume, addresses this drawback.
Given that the interaction length of
fission-energy neutrons in organic materials is a few cm, the detector's principle 
relies on the ability to resolve two proton recoils in a contiguous scintillator 
volume at spatial and temporal separations of order \cm{1} and \genunit{1}{ns} 
respectively.
This reconstruction of the neutron 
interaction locations must be based on the arrival time and position
of the scintillation photons at the boundaries of the active volume.
While traditional photomultiplier tubes (PMTs) cannot provide the necessary
spatial and temporal resolution, recent advances in photodetector (PD) technology 
have made the approach possible~\cite{Krizan_2014}. Photodetectors 
based on micro-channel plate (MCP) electron multipliers~\cite{Wiza_1979} 
inherently provide the spatial and temporal photon detection resolution that 
makes them attractive for single-volume event reconstruction. These MCP-PMTs have 
found application in a wide variety of areas ranging from medical
imaging~\cite{Kim_2012} to neutrino detection~\cite{Li_2016a,Back_2017}.
Commercially available photodetectors have \genunit{2-6}{mm} two-dimensional
pixels and sub-ns timing~\cite{planacon_datasheet}, while models 
under development will potentially provide large area 
photocathode coverage at relatively low cost~\cite{Adams_2015}.
In this work, we investigate
the feasibility of the SVSC concept using Monte Carlo simulation studies
of a system with PD performance motivated by available MCP-PMTs.

Throughout this paper, the term ``single-volume'' means that the active 
detection volume is compact and essentially contiguous, rather than spread 
out spatially in discrete well separated volumes (cells, planes, etc.).
We refer to an individual neutron scatter, most usefully neutron-proton 
elastic scattering, as an ``interaction'', and to a neutron history, which 
may include multiple interactions, as an ``event''. In an 
experimental context, an event can mean the data collected by a single 
trigger, which could include information from zero (noise), one, or more
(pileup) particles. ``Event reconstruction'' is an algorithm that uses the 
observable optical photon information to determine the locations and times of 
neutron interactions in the event. Many such reconstructed events can then be 
combined to reconstruct a neutron image and energy spectrum.

This article is structured as follows. We begin in \secref{s:svscGeneric} 
with the general motivation and design-agnostic arguments for a single-volume 
neutron scatter camera, as well as a survey of several possible design 
classes. In \secref{s:svscDesign}, we describe the ``direct reconstruction'' 
SVSC concept as simulated in this paper, emphasizing how the design choices 
are enabled by existing technology. The data processing and reconstruction 
steps required by this concept are also detailed in \secref{s:svscDesign}. 
The rest of the paper is dedicated to a demonstration of the feasibility of 
the SVSC concept based on Monte Carlo simulation data. Details of the 
simulated model and the corresponding physical behavior of the system 
directly obtained from the simulation are presented in \secref{s:simulation}. 
The results of the data processing stage applied to the simulated data, which 
will cover the event reconstruction and the source spectral and image 
reconstruction, are presented in \secref{s:results} together with an analysis 
of the instrument's limitations.

\section{General considerations for a single-volume neutron camera}
\label{s:svscGeneric}

We begin with a discussion of the motivation for and approaches to single-
volume neutron imaging, without reference to a specific concrete detector 
system design.

\subsection{Advantages of single-volume neutron imagers}
\label{ss:svscMotivation}

The motivation for research into a single-volume double-scatter neutron 
imager rests on the potential for significant improvements over cell-based 
scatter camera designs in two important aspects: imaging efficiency and 
compactness. 

A single-volume camera addresses two of the important efficiency limitations of 
the NSC: the requirement of no more than one neutron scatter in the first 
detector element (in order for a one-to-one relationship between fractional 
energy deposited and scattering angle to be valid), and the geometrical 
efficiency for the scattered neutron to pass through a second detector 
element. Assuming the ability to resolve scattering events at \cm{2} 
separation in the single volume, an estimate of the improvement in efficiency 
for double-scatter neutron detection, relative to the NSC of~\cite{Brennan_2011},
is about an order of magnitude.

To obtain this estimate, an MCNP~\cite{Pozzi_2003} simulation was performed: a 
fission neutron pencil beam was fired at the center of one of the middle 
detectors in the conventional NSC front plane, and at the center of a
$\cm{20}\times\cm{20}\times\cm{20}$ volume of scintillator. The NSC was
simulated with a standard plane separation of \cm{40}.
The fraction of events satisfying simple requirements were tallied:
first, that two interactions occur---as a function of the distance between the
first two neutron scatter interactions for
the single-volume system, and with a front and rear plane requirement for the
current NSC; second, that 
both interactions were hydrogen scatters (carbon recoils are typically too 
quenched to observe); and third, that both interactions deposited at least 
\kev{200}, an optimistic threshold.
In \fig{f:svscvnsc}, the results are shown for the single-volume system as a
function of the distance between the first two interactions. The corresponding
fractions for the cell-based NSC are 0.026 for the first two interactions in the
front and rear planes; 0.013 for both scatters producing proton recoils; and
0.008 for both protons having at least \kev{200}.
The SVSC yields a significant improvement over the conventional NSC
in potential efficiency when the minimum separation between interactions is 
low. For a \cm{2} minimum interaction spacing, the number of potentially 
detectable events is slightly over an order of magnitude higher in the SVSC 
than in the current NSC. Even at \cm{10} minimum interaction spacing, the 
fraction of usable neutrons is comparable to the current NSC. Further study 
will be needed to understand the relative power of events with small 
separation in a SVSC, since the energy and imaging resolutions will degrade 
with smaller separation distances.

\begin{cfigure}
\includegraphics[width=0.95\columnwidth]{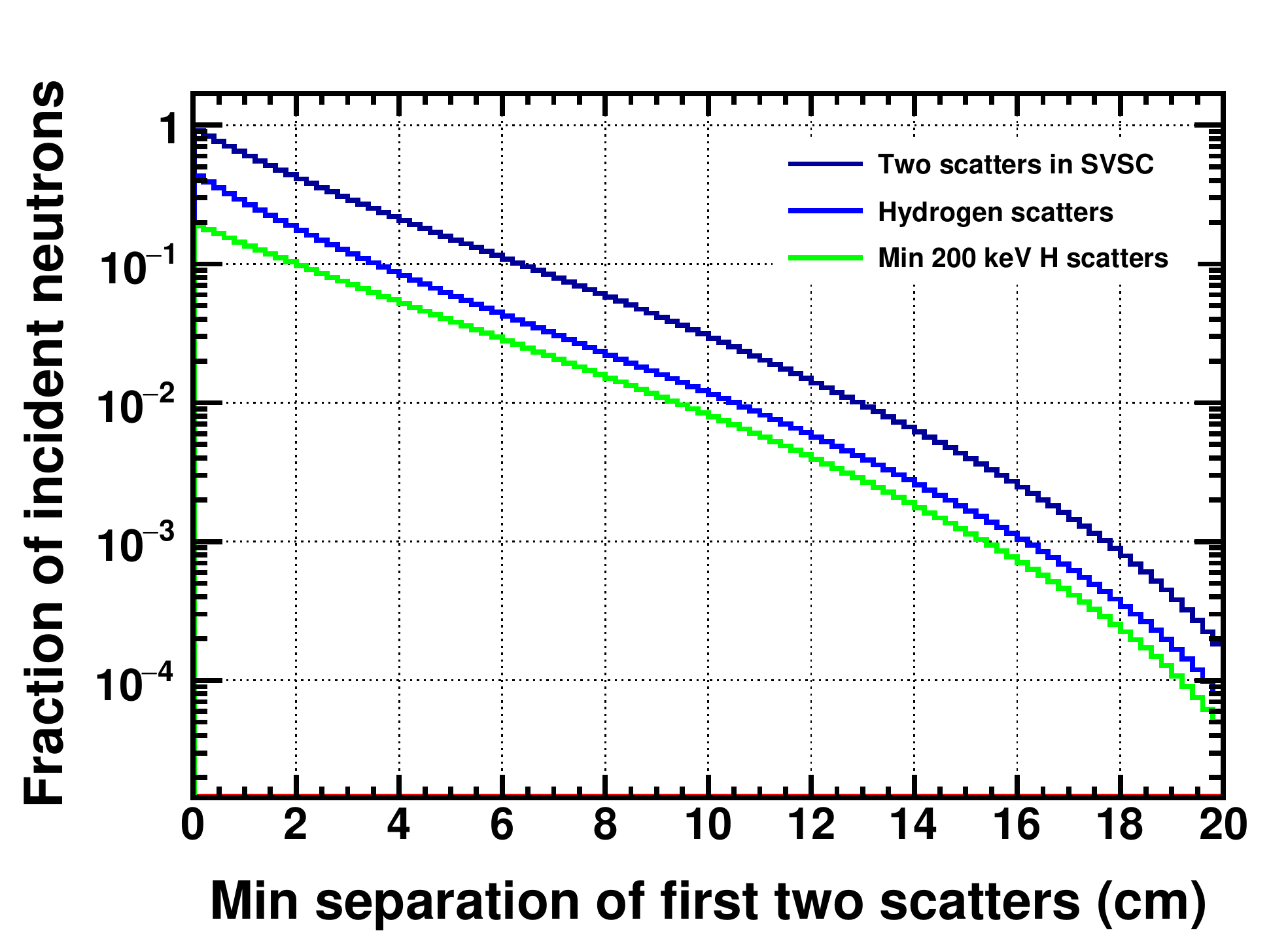}
\caption{Potential efficiency for the proposed single-volume scatter camera.
Details of the MCNP simulation are given in the text. No experimental
resolution effect or detection efficiency is included, beyond the requirements
listed. The values are given as a function of the distance separating the 
first two interactions in the volume; the ability to distinguish interactions 
at smaller separations increases the potential efficiency of the detector.}
\label{f:svscvnsc}
\end{cfigure}

Additionally, the overall detector footprint is more compact and lighter: 
about $\cm{40}\times\cm{40}\times\cm{20}$, with expected weight about
\genunit{10-20}{kg} for the same scintillator volume as the 32-element SNL 
NSC, which 
takes up a volume of approximately \mcub{1} and weighs more than
\genunit{200}{kg}. Easier transport and deployment is a key benefit for some 
applications. Moreover, a more compact detector allows to get closer to the 
source, which in turn increases detection rate according to the inverse 
square of the distance to the source, and improves spatial resolution at the 
object for a given angular resolution.

\subsubsection{Four single-volume imager concepts}
\label{sss:fourApproaches}
Single-volume neutron imaging is predicated on the ability to use the 
information encoded in the optical photons emitted by the scintillator to 
reconstruct the scattering history of the neutron in the active volume. Even 
if it is granted that from an information theory perspective there is in 
fact sufficient information retained by the optical photons to do this, it is 
still one of the key technical challenges of this research to develop 
detector designs and algorithms that accurately (and tractably) do it.
Four general methods that have been considered for building a single-volume 
neutron imager are as follows.
\begin{description}
\item[Direct reconstruction]
This method is conceptually the simplest and yet technically the most
challenging. The scintillator is a single monolithic volume, and 
optical photons are detected at the boundaries of the volume. The neutron 
interactions are reconstructed based on the information in the detected 
photon positions and times, accounting for the isotropic photon emission from 
the scintillator, the scintillator pulse shape, photon speed and attenuation 
in the scintillator, photodetector efficiency, etc. This paper documents studies
of this concept.
\item[Optical segmentation]
The ``single volume'' of scintillator is formed from multiple pillars of 
scintillator, separated by reflective boundary layers. The goal is to trap 
the optical photons from an interaction in one pillar, reading them out from 
both ends. For each interaction, the location along the long dimension of 
the pillar is determined by the amplitude ratio and/or time difference of the 
light observed at each end, while the transverse location is determined by 
which pillar the light is observed in. When two interactions occur in 
different pillars, their locations and times can be independently determined, 
constituting an event reconstruction. Simulation studies toward a realization
of this concept are detailed in~\cite{Weinfurther_2017}.
\item[Optical coded aperture]
This method, like direct reconstruction, uses a single uninterrupted 
scintillator volume, but instead of directly detecting and reading out the 
optical photons on the boundary of the volume, they pass through a coded 
aperture on each side. This has the advantage of introducing a high-frequency 
spatial component to the detected photon distribution, which eases the 
spatial reconstruction of interaction locations. Disadvantages include the 
loss of a significant fraction (typically 50\%) of the photons to the mask, 
and the presence of light guide material, which is necessary to allow the 
photons to spread before and after the coded mask, but also results in an 
inactive scattering/attenuating layer that neutrons must penetrate before 
reaching the active volume. A similar approach has been studied for gamma
Compton imaging~\cite{Ziock_2015}.
\item[Optical lattice]
This is similar to the optical segmentation approach, but aims to create 
virtual ``pillars'' in all three dimensions. Small cubes of scintillator are 
separated by very thin air gaps in a three-dimensional array. The
scintillator-air boundaries create total internal reflection (TIR) for 
photons hitting 
them at larger than the critical angle. Some photons experience TIR in each 
of the three dimensions of the lattice, and are therefore confined to a 
single ``pillar'' or row of cubes. Light therefore arrives as a bright spot on
all six sides of the array, not just two. This approach has been studied for
neutrino detection~\cite{Lane_2015}.
\end{description}

It should be noted that the reconstruction uncertainties, and consequently 
the directional and spectral resolving power, of each event in single-volume 
designs depend significantly on the design details, component performance
(scintillator time profile, photodetector efficiency and resolution, etc.), and 
reconstruction algorithms used. On the one hand, a single-volume design will 
be able to determine the location and time of neutron interactions with 
better resolution than the NSC. On the other hand, we expect reduced 
precision in the measurements of deposited energy when disentangling the 
light output from multiple interactions, and the short lever arm between the 
two neutron interactions also decreases imaging resolution with 
respect to the cell-based design. In the remainder of this paper, we focus on 
further exploring the direct reconstruction technique.

\section{Single-volume scatter camera design}
\label{s:svscDesign}

We now proceed to specify a concrete detector system design and reconstruction
algorithm for full simulation studies of a single-volume imager.

\subsection{SVSC system as simulated}
\label{ss:detectionSystem}

The direct reconstruction SVSC design concept, schematically shown in
\fig{f:svscCartoon}, consists of an organic scintillator volume with each side 
optically coupled to and fully covered by fast-timing high-gain MCP-PMTs. 
These photodetectors play the central role of registering the arrival time 
and position of the isotropically emitted photons produced by particle 
interactions in the scintillator volume. As already mentioned, current
MCP-PMT technology, either available or under development, can provide photon 
timing resolution on the order of \genunit{100}{ps} and mm-scale spatial 
photon detection resolution. The simulation studies presented in
\secref{s:simulation} show the impact on the event reconstruction due to such 
finite MCP-PMT resolution values by comparing results with idealized 
photodetectors able to provide exact time and spatial photon coordinates, and 
demonstrate 
the detector's spectral and image reconstruction capabilities even with 
existing MCP-PMT technology. Another enabling consideration of the SVSC 
concept is ability of high speed waveform sampling required to take advantage 
of the intrinsic time resolution of MCP-PMTs, and especially to disentangle 
multiple photons arriving in one photodetector pixel within a timescale of 
nanoseconds. This is possible using switched capacitor array sampling, such 
as with the Domino Ring Sampler chips, which provide up to \genunit{5}{GS/s} 
sampling frequency~\cite{Ritt_2010,Bitossi_2016,Oberla_2014}

\begin{cfigure}
\includegraphics[width=0.99\columnwidth]{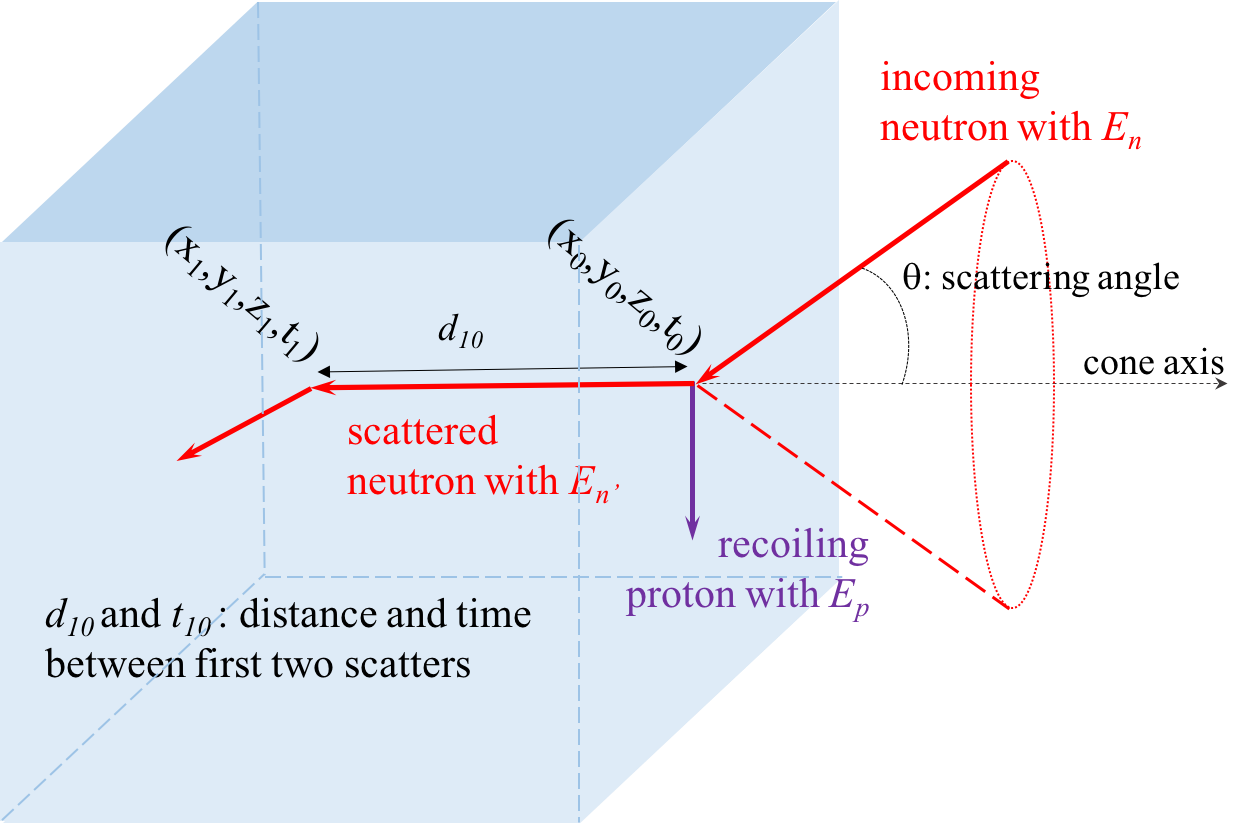}
\caption{Sketch of a neutron event in the SVSC. The neutron interacts twice 
in the SVSC scintillator volume. If the position and time of the two 
interactions can be reconstructed using the optical photons emitted in the 
scintillator, then the incoming neutron energy and direction (up to a conical 
ambiguity) can be determined through the kinematic properties of
neutron-proton elastic scattering.}
\label{f:svscCartoon}
\end{cfigure}

When selecting the active material type, fast scintillation decay time 
represents the main driver, even more important than the ability to 
discriminate neutrons and gammas via pulse shape discrimination (PSD). 
Narrow 
scintillation pulses allow for better event timing reconstruction, especially 
when reconstructing events with several neutron interactions.
Although PSD is desirable for this system, it may be superfluous 
since we already require the ability to reconstruct the time of flight of the 
detected particle, which provides neutron/gamma discrimination. 
High light output is also desired, as photon statistics is one limiting factor
in the event reconstruction described below, although we note that the readout
and signal processing are simplified if the photon
occupancy per MCP-PMT pixel is low. Here we simulate EJ-232Q
~\cite{ej232q_datasheet}, a quenched plastic scintillator with 
pulse width of less than \genunit{400}{ps} but only 19\% anthracene light
output, a choice that 
emphasizes the pulse width over photostatistics. The scintillator size is 
set to a \genunit{10}{cm} cube in order to guarantee a high probability of 
double neutron scatters, in accordance with the interaction length for
fission-energy neutrons. In general, the choice of active volume size will be 
dictated by a variety of factors, like detection rate and efficiency 
requirements, hardware constraints and application-specific trade-offs, which 
go beyond the scope of this paper.

\subsection{Reconstruction algorithms}
\label{ss:dataProcessing}
Data processing and analysis are essential elements in the design and 
operation of the SVSC. They are accomplished in three separable analysis stages: 
single-photon isolation in each individual pixel, neutron double-scatter 
event reconstruction, and the final image and spectrum reconstruction.

Single-photon isolation is the process of assigning an arrival time and 
location to each detected photon using the digitized raw traces from each
MCP-PMT channel (or pixel). The specifics of this part of the analysis are highly
coupled to the experimental behavior of the photodetector and electronics.
For this simulation study, since the goal is to determine feasibility of the
concept, we make the optimistic assumption that all detected
photons can be individually identified.

For event reconstruction, we employ the direct 
reconstruction of particle interactions using the arrival position 
and time of isotropically emitted photons at the scintillator surface.
Multiple methods have been explored; here
we use the unbinned maximum likelihood (ML) 
technique. The likelihood function is maximized for a given event when the 
fixed set of observations (the list of detected photon coordinates and times) 
has the highest probability of occurring as the event parameters (the list of 
neutron interaction locations and times) are varied. Put another way, the ML 
algorithm finds the neutron trajectory that would be most likely to produce 
the observed data. We use the extended maximum likelihood~\cite{Barlow_1990},
\begin{equation}
\label{eq:likelihood}
\mathcal{L}=\frac{\mathrm{e}^{-\mu}\mu^{n}}{n!}\prod_{i=1}^{n}\sum_{j=1}^{N}
\frac{\mu_{j}}{\mu}P_{j}(\vec{x_{i}}),
\end{equation}
where $\mu_{j}$ is the number of photons detected from neutron interaction $j$
and $\mu=\sum\mu_{j}$, $n$ is the number of detected photons, $N$ is the 
number of neutron interactions assumed, $\vec{x_{i}}$ represents the detected 
position and time of photon $i$, and $P_{j}(\vec{x_{i}})$ is the probability 
to observe photon $i$ from interaction $j$. Each photon could have come from 
any of the neutron interactions, so the probability to observe each photon is 
summed across interactions; each photon is an independent observation, so the 
photon probabilities multiply. Looking more closely at the probability 
function, we have
\begin{equation}
P_{j}(\vec{x_{i}})=\frac{\cos\phi_{ij}}{4\mathrm{\pi}d_{ij}^{2}}\cdot
\mathrm{e}^{\frac{-d_{ij}}{\lambda}}\cdot f(t_{i}-t_{j}-d_{ij}/c_{p}).
\end{equation}
The first term accounts for solid angle between the location of emission
(interaction $j$) and the location of detection for photon $i$. The angle 
between the line connecting the two locations and the normal of the 
photodetector surface is $\cos\phi_{ij}$, and $d_{ij}$ is the distance 
between the two locations. The second term accounts for attenuation in the 
scintillator material, where $\lambda$ is the attenuation length. In the 
third term, $f(t)$ is the pulse shape of the scintillator, where the time 
difference between the photon emission and detection has been corrected for 
the flight time of the photon ($c_{p}$ is the speed of light in the medium).
The likelihood maximization itself is performed using the SIMPLEX and
MIGRAD algorithms as implemented by the MINUIT package in the
ROOT software~\cite{James:1994vla}.

This likelihood maximization results in an estimate of the position, time, and
intensity of each neutron interaction, and an associated variance-covariance
matrix (via inverting the Hessian matrix at the log-likelihood maximum).
From these fitted quantities, we then derive \dsep, \tsep, and the cone axis
direction, as shown in \fig{f:svscCartoon}, and their uncertainties. Finally,
the kinematic equations \eqnrange{eq:kin1}{eq:kin3} are used to determine the
ultimate quantities of interest, namely \En and the cone opening angle $\theta$.
In all cases, the fit errors are propagated to the derived results via the
standard linearized approximation, including correlations.

Finally, image reconstruction is performed from many such neutron events 
using a Maximum Likelihood Estimation Maximization (MLEM) algorithm
~\cite{Shepp_1982}. We use a list-mode implementation of MLEM, in 
which a simplistic system response is calculated for each observed event, 
under the assumption that the reconstructed cone axis and opening angle are
accurate within uncertainties.

\section{Simulation}
\label{s:simulation}
A Monte Carlo model of the SVSC detector system was constructed using
Geant v4.10.01.p02~\cite{Agostinelli_2003}.
The scintillator material as described in
\secref{ss:detectionSystem} is centered at the coordinate origin. A
\genunit{1}{mm} 
thick acrylic light guide representing an optical coupling encloses the 
scintillator, which is then completely surrounded by a \genunit{1}{mm} quartz 
layer of very short optical absorption length representing the MCP-PMT. An isotropic 
point-like \Cftft neutron source is placed at \meters{1} from the origin 
along the $y$ axis. 
 
The simulation uses the physical attributes listed in
\tab{t:simAttributes}. Neutron interactions are modeled 
by the Geant4 NeutronHP cross-section package, which includes high precision 
hadronic models for energies below \mev{20}, appropriate for fission-energy
neutrons. Scintillation photons from 
electron recoils are generated according to the manufacturer-reported 
scintillation efficiency for EJ-232Q with 0.5\% benzophenone concentration. 
Due to lack of experimental data on proton quenching in this scintillator,
we use the energy-dependent proton light 
output measured for a similar plastic scintillator in~\cite{Pozzi_2004}, and 
we model the carbon recoil light output according to~\cite{Batchelor_1961}. The 
photon generation time profile follows the manufacturer-reported exponential 
rise and decays times, shown in \tab{t:simAttributes}, with a reported
scintillation pulse width of \genunit{360}{ps}.
The optical interfaces between the scintillator, the light guide and the photodetector 
are assumed to be perfectly smooth, and photons are either reflected or transmitted but are 
not absorbed in the materials interface. A wavelength-dependent refraction index is used 
for the acrylic light guide~\cite{Kasarova_2007}. 
Optical photon detections per event are recorded over a \genunit{100}{ns} window.
In order to account for the photocathode 
quantum efficiency, photons absorbed in the photodetector volume are randomly 
saved with 25\% efficiency.

\begin{table}
\centering
\caption{Numerical values of various parameters used in the GEANT4 simulation 
of the SVSC concept. Component abbreviations: SC refers to the scintillator; 
LG refers to the light guide; PD refers to the photodetector.}
\label{t:simAttributes}
\begin{tabular}{ll}
\hline
Attribute	& Value \\
\hline
SC cube size & \genunit{100}{mm} \\
SC Efficiency & \genunit{2900}{ph/MeVee} \\
SC pulse rise time	& \genunit{0.11}{ns} \\
SC pulse decay time & \genunit{0.7}{ns} \\
SC absorption length & \genunit{8}{cm} \\
SC refraction index & 1.58 \\
LG refraction index & 1.49--1.59\\ 
PD quantum efficiency & 0.25 \\
PD absorption length & \genunit{10^{-3}}{mm} \\
PD refraction index & 1.57 \\
PD time spread & \genunit{0.1}{ns} \\
PD pixel size & \genunit{5.9}{mm} \\
\hline
\end{tabular}
\end{table}

The left plot of \fig{f:truthResults1} shows the histograms of the number of 
neutron elastic collisions within the scintillator volume, normalized by the 
number of primary neutrons that interact in the detector.
When a \kev{300} threshold is imposed on the 
neutron deposited energy---which corresponds to about \kevee{40} of proton 
light output in EJ-232Q or, equivalently, to only about 30 detected
photons---the number of above-threshold neutron collisions per event is 
reduced drastically. This indicates that the transfer of most of the 
neutron's energy to the scintillator is usually done in few elastic scatters.

\begin{cfigure1c}
\includegraphics[width=0.32\textwidth]{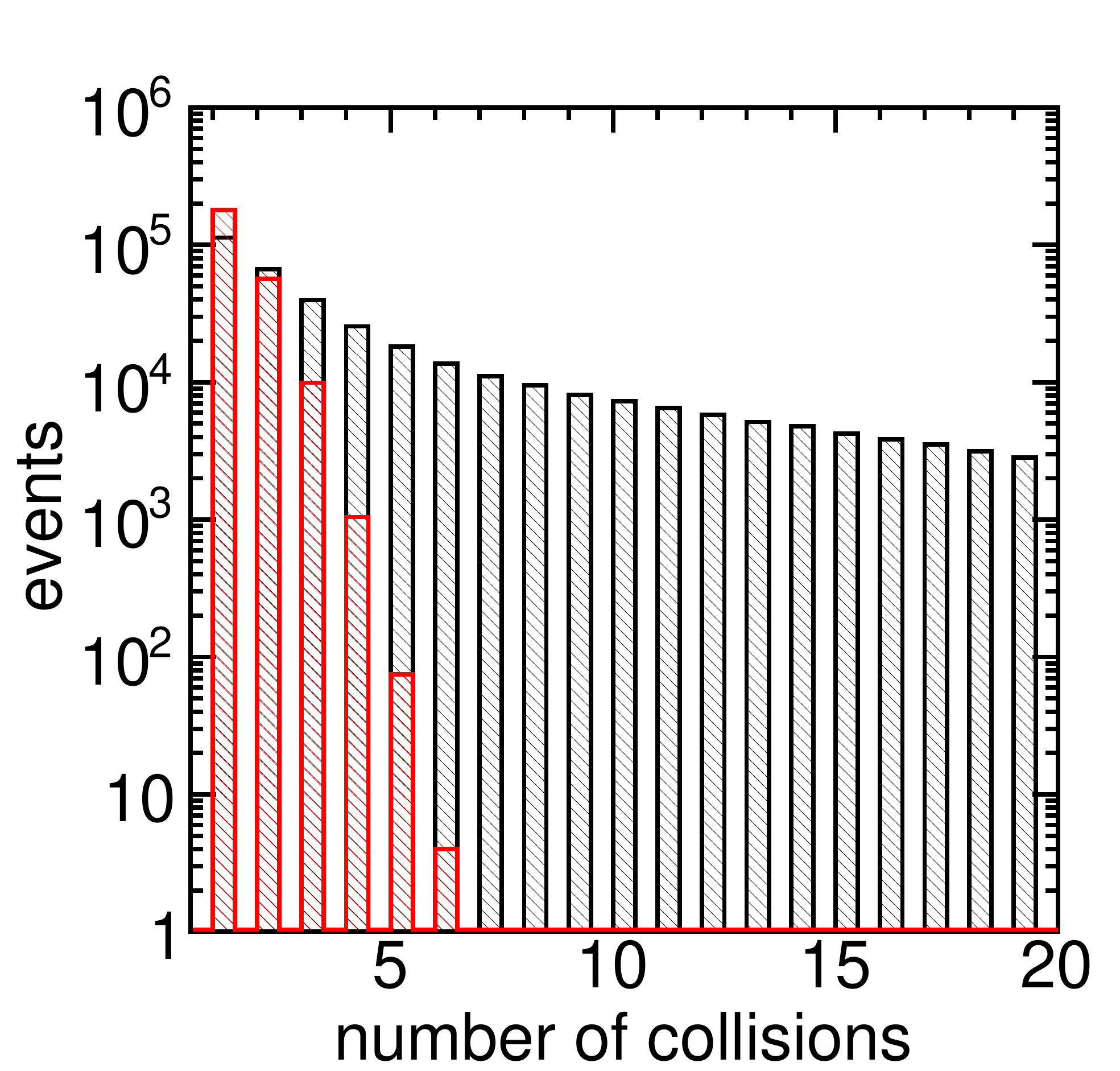}
\includegraphics[width=0.32\textwidth]{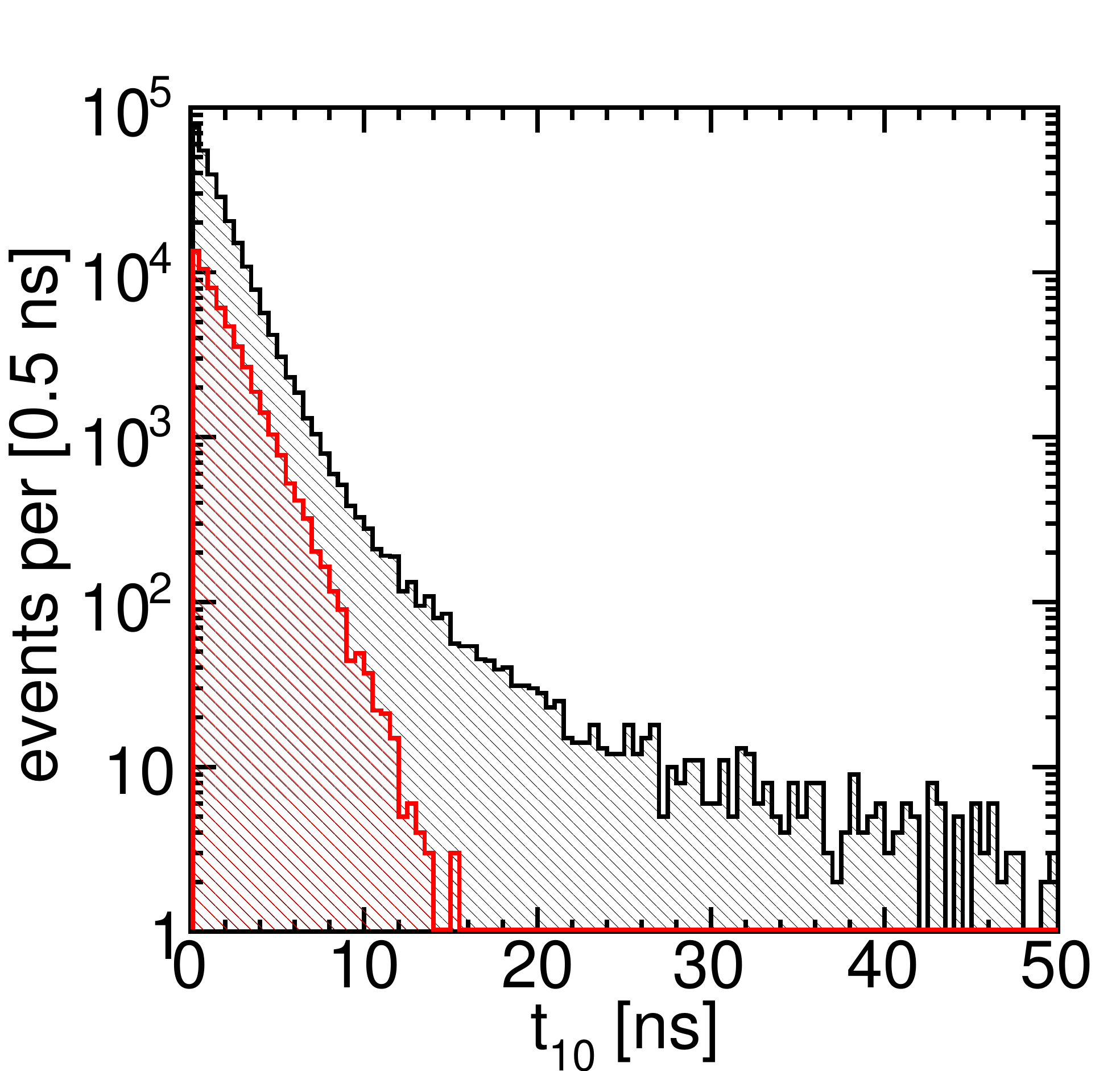}
\includegraphics[width=0.32\textwidth]{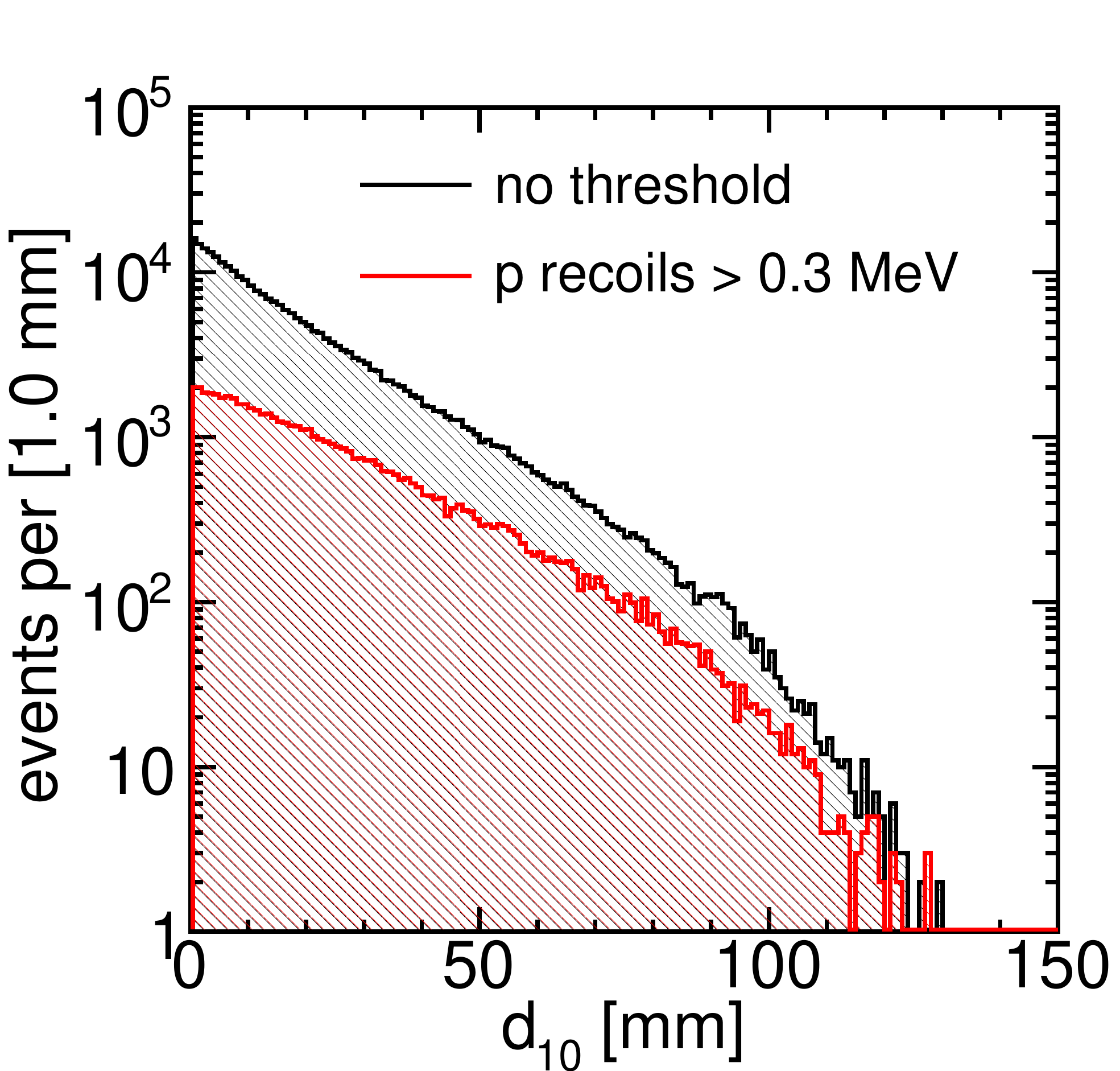}
\caption{Truth-level results from the GEANT4 simulation of a \Cftft source. 
Left: The number of neutron elastic collisions.
Middle: The time difference between the first two neutron interactions,
for neutrons interacting at least twice in the scintillator.
Right: The distance between the first two neutron interactions,
for neutrons interacting at least twice in the scintillator.
No threshold on the deposited energy was imposed on the neutron interactions 
populating the histograms in black, while the histograms in red only include 
interactions creating proton recoils that deposit more than \kev{300}. Moreover,  
only events where the first two proton recoils are above the threshold are included
in the \tsep and \dsep red histograms. 
}
\label{f:truthResults1}
\end{cfigure1c}

The middle and right histograms of \Fig{f:truthResults1} show the histograms
of the time and distance between the first two neutron collisions.
These quantities, along with the energy transferred in the first
collision, contain the information from which the 
neutron's incoming energy and direction can be extracted. For zero 
threshold, the mean distance 
between the first two neutron collisions is about \cm{1.8}, increasing to
\cm{2.4} for collisions above \kev{300}, and the mean time remains
\genunit{\sim 1.7}{ns} in both threshold cases. 

The arrival time and spatial coordinates of the individual optical photons 
from all neutron collisions, regardless of deposited energy, constitute the 
input data for the event reconstruction algorithm described above.
In order to include the effect of finite timing resolution in
the MCP-PMT and electronics, the photon arrival time was Gaussian 
smeared with standard deviation equal to \genunit{0.1}{ns}, which is informed
by the transit time spread in existing detectors and bandwidth of available
electronics~\cite{planacon_datasheet,Grigoryev_2016}. The 
photon two-dimensional coordinates along each photodetector plane---e.g., the 
$(x,y)$ coordinates of the photodetector at $z=\genunit{50}{mm}$---were also 
Gaussian smeared. For the coordinates smearing, we use a standard deviation 
equal to $(\genunit{5.9}{mm})/\sqrt{12}=\genunit{1.7}{mm}$, which matches the 
standard deviation of a rectangular distribution with width \genunit{5.9}{mm} 
representing the pixel size of the Planacon XP85012. 

In the following two ways, the current study takes advantage of information
that is known in simulated events but not in real data, and therefore
potentially over-estimates the system performance.
First, we constrain the total number of neutron collisions, $N$, to be the simulated 
number of neutron scatters depositing more than \kev{300}.
In real data, it should be possible to estimate the number of energetic collisions 
in the scintillator volume from the spatially integrated photon detection time profile 
combined with the time-sliced spatial distributions of photon detection locations.
We note that there is no obviously ``correct'' value for $N$: 
counting real but low-energy collisions into $N$ adds irrelevant degrees of 
freedom to the MINUIT minimization, increasing the probability of incorrect
convergence of \eqnref{eq:likelihood}. A process of iterative 
minimizations with $N$ ranging over most likely values according to the 
observed signals could be part of the event reconstruction algorithm when 
working with real data.

Second, in this study, small deviations from the interactions' actual
location coordinates (\genunit{0.5}{mm}), time (\genunit{0.1}{ns}), 
and number of emitted photons (50) are used to set the 
initial guesses in the minimization algorithm. (Events for which the minimization
algorithm does not move from these initial parameter values are rejected as
failed fits.)
In the limit of a globally convex likelihood function, this procedure does not
produce a bias, as the fit will converge to the same global minimum regardless
of the initial position. However, in the case that the function has multiple
local minima, this procedure could bias the results by over-estimating the
accuracy of the initial guess that could be achieved.
Again, in real data, initial guesses could be extracted from rough analysis of
the photon detections, but we leave the development and demonstration
of such an algorithm to future work. 

Idealizations in our simulated model could play a more 
relevant role in the reconstruction algorithm success.
Complete wavelength dependence of optical processes is not included.
The optical interfaces 
between the scintillator, the light guide and the photocathode are assumed to 
be perfectly polished surfaces with nearly matching indexes of refraction, 
and thus, most simulated photons are transmitted and absorbed in the 
photocathode. However, a realistic detector will likely suffer from increased 
photon scattering in the optical surfaces due to imperfect optical coupling 
and surface roughness. Any scattered photons that are later detected by the 
photodetectors will not satisfy the reconstruction's assumption of being 
directly emitted from interaction locations, and thus, will affect the event 
reconstruction success. Additionally, the detected photon count of a real 
detector is expected to be smaller than that of our simulated model due to 
incomplete photocathode coverage and other unmodeled optical absorption
effects. 
Furthermore, we assume in simulation that each 
photon can be individually resolved and accurate arrival time can be 
assigned, but the overlapping of their corresponding single-photoelectron 
waveforms, already discussed in \secref{s:svscDesign}, will affect the photon 
count and timing per photodetector pixel. Finally, here we simulate and
reconstruct only neutron events. In reality, some gamma interactions will
be misidentified as neutrons and contribute erroneously to a neutron image.

Due to the assumptions and limitations of the simulation model described above,
the event reconstruction 
and the image reconstruction results presented in the next section
should be understood as an upper limit on the
performance of the SVSC detector concept. Put another way, we aim here to establish feasibility
of the SVSC design, rather than to accurately predict performance of a system.

\section{Simulation results}
\label{s:results}
In order to quantify the event reconstruction performance, we define the
``true'' variables in a given event as the positions, times, and intensities
of the first two n-p scatters in the simulation, as well as the higher-level
quantities (e.g.\ \dsep or $\theta$) calculated from those. Note that in some
classes of events, such as those with an initial carbon scatter before two
hydrogen scatters, the ``true'' reconstructed cone may not include the incident
direction since the kinematic assumptions are not valid. But the goal of the
event reconstruction is to determine the attributes of the proton recoils, so
we evaluate it on that basis.

For each variable, we construct resolution plots, i.e.\ histograms of the
difference between the reconstructed and the true variables; these are plotted
in \figrange{f:reco1}{f:comp1} and will be discussed below in turn. The
differences are denoted with the symbol $\Delta$ followed by the variable
symbols defined in \fig{f:svscCartoon}. In these plots, the photon arrival time
$T$ and coordinates $(X,Y,Z)$ that entered the reconstruction algorithms have
been smeared as described in the previous section, and only events for which
the likelihood maximization converged are included. Each histogram is plotted
with a fitted Gaussian function, whose mean ($\mu$) and width ($\sigma$) are
given in the plot as well as in \tab{t:perf}.

\begin{table*}
\centering
\caption{The event reconstruction resolutions are summarized in this table.
The mean $\mu$ and width $\sigma$ of Gaussian fits to the 
various reconstructed quantities are given; this represents the bias and 
resolution of the reconstruction for this simulated system and \Cftft source. 
Refer to \fig{f:svscCartoon} for definitions of the variables. Results are 
presented for reconstructions using both smeared (representing
experimental resolutions) and the exact (as obtained from the simulation) 
photon positions and times. See text for discussion of the $\Delta\alpha$
distribution.}
\label{t:perf}
\begin{tabular}{lSSSS}
\hline
 & \multicolumn{2}{c}{Smeared} & \multicolumn{2}{c}{Exact} \\
 & \multicolumn{1}{c}{$\mu$} & \multicolumn{1}{c}{$\sigma$} &
   \multicolumn{1}{c}{$\mu$} & \multicolumn{1}{c}{$\sigma$} \\
\hline
$\Delta x_{0}$ (mm)    & 0.0096 & 3.2 & 0.0028 & 2.9\\ 
$\Delta y_{0}$ (mm)    & 0.88 & 3.3 & 0.90 & 2.8\\ 
$\Delta z_{0}$ (mm)    & 0.0057 & 3.2 & -0.0060 & 2.9\\ 
$\Delta t_{0}$ (ns)    & 0.026 & 0.081 & -0.038 & 0.019\\ 
$\Delta n_{0}$         & -46 & 43 & -39 & 31\\ 
$\Delta x_{1}$ (mm)    & -0.023 & 5.4 & -0.12 & 4.6\\ 
$\Delta y_{1}$ (mm)    & 0.35 & 5.5 & 0.15 & 4.6\\ 
$\Delta z_{1}$ (mm)    & 0.029 & 5.3 & -0.043 & 4.5\\ 
$\Delta t_{1}$ (ns)    & -0.017 & 0.15 & -0.039 & 0.034\\ 
$\Delta n_{1}$         & -28 & 30 & -27 & 24\\ 
$\Delta d_{10}$ (mm)   & 5.6 & 7.1 & 3.9 & 5.8\\ 
$\Delta t_{10}$ (ns)   & -0.024 & 0.17 & 0.00081 & 0.047\\ 
$\Delta\Ep$ (\mevnoarg)& -0.33 & 0.28 & -0.31 & 0.23\\ 
$\Delta\En$ (\mevnoarg)& -0.24 & 0.45 & -0.25 & 0.45\\ 
$\Delta\theta$         &-0.12 & 0.10 & -0.12 & 0.096\\ 
$\Delta\alpha$         &      & 0.06 &       & 0.07\\
\hline
\end{tabular}
\end{table*}

We observe first that the distributions are not well described by a single
Gaussian. We attribute this to two effects. First, each distribution includes
a large diversity of different types of events, with variability in the
interactions' distance to the photodetectors, in the distance between
interactions, in the energy allocation among the interactions, in the total
number of photon-emitting interactions, etc. As such, even if event
reconstruction resulted in accurate and normal uncertainties for each type of
event, these histograms would not necessarily be Gaussian in shape since they
are effectively composed of multiple distributions, each with its own standard
deviation. Second, some events are reconstructed with an erroneous topology.
For example, if the first two n-p scatters are very close in space and time,
they may be reconstructed as a single interaction, and the third scatter
reconstructed as the second. This second type of error will lead to very long
non-Gaussian tails in the distributions.

Still, the widths of the histograms, expressed in
terms of the standard deviations $\sigma$ of the corresponding Gaussian fits,
provide a measure of the resolution achievable with the event reconstruction 
algorithm, while the Gaussian means $\mu$ resulting in values significantly 
shifted from zero represent biases in the reconstruction. The first pair of  
columns of \tab{t:perf} list the values of $\sigma$ and $\mu$ for all plotted 
variables where the smeared time $T$ and coordinates $(X,Y,Z)$ 
were input to the algorithm, while the second pair gives the results when 
the exact $T$ and $(X,Y,Z)$, before the Gaussian smearing, were used.  
With a similar column breakdown,
\Tab{t:RecoSuccessRate} contains efficiency results, including the fraction of
incident neutrons that contain a given number of proton recoils above a
\kev{300} threshold, and the fraction of those for which the reconstruction
is ``successful'', defined as the fraction of events for which
the likelihood maximization converges, with the added restriction that the 
reconstructed locations and times of the first two interactions are 
within \genunit{10}{mm} and \genunit{2}{ns} of the true values, respectively.
The latter condition removes some, but not all, of the events reconstructed
with an erroneous topology. As in \tab{t:perf}, these values are given for the
reconstructions with and without resolution smearing.

\begin{table*}
\centering
\caption{Reconstruction efficiency is summarized in this table.
For each given number of proton scatters above the \kev{300} recoil energy
threshold, we show the fraction of incident neutrons in this category, and the
fraction of such events that are successfully reconstructed (as defined in the
text), with and without photon detection resolution smearing.}
\label{t:RecoSuccessRate}
\begin{tabular}{cSSS}
\hline
\# proton recoils &
 \multicolumn{1}{c}{Fraction of incident} &
 \multicolumn{2}{c}{Reconstruction success rate (\%)} \\
$\geq\kev{300}$ &
 \multicolumn{1}{c}{fission neutrons (\%)} &
 \multicolumn{1}{c}{Smeared $\vec{x_{i}}$} &
 \multicolumn{1}{c}{Exact $\vec{x_{i}}$} \\
\hline
2 & 11.4 & 49 & 62 \\ 
3 &  2.1 & 34 & 49 \\ 
4 &  0.2 & 24 & 37 \\ 
\hline
\end{tabular}
\end{table*}

\Figrange{f:reco1}{f:reco3} show the $\Delta$ histograms for the locations, 
times, and detected photon count of the first and second above-threshold proton 
recoils: (\Dxzero, \Dyzero, \Dzzero, \Dtzero, \Dnzero) and (\Dxone, \Dyone, 
\Dzone, \Dtone, \Dnone), respectively. These histograms represent the resolutions 
of the individual interaction variables directly obtained from the event 
reconstruction algorithm. The reconstruction resolution for the first interaction
is about \genunit{3.2}{mm} in each spatial dimension and \genunit{81}{ps} in
time. The reconstruction of the second interaction is somehwat less accurate,
as indicated by larger widths of the corresponding histograms. One possible 
cause is that below-threshold collisions tend to occur later in the neutron 
trajectory as its kinetic energy decreases, and photons emitted from those 
below-threshold collisions will be more likely assigned to the second 
reconstructed interaction. Added to this is the fact that the energy 
deposited in the second interaction is on average smaller than the 
energy deposited in the first interaction, producing a smaller second 
photon pulse which provides relatively less information to the reconstruction
algorithm.

The \Dnzero and \Dnone plots show a more significant bias for these quantities,
on the order of their resolution. The reconstructed $n_i$ quantities, defined as
the number of photons detected from interaction $i$, are simply the $\mu_i$
from the maximized likelihood (\eqnref{eq:likelihood}). The true quantities
are the \emph{expected} number of photons generated in that interaction (i.e.\
before Poisson fluctuation) multiplied by the photodetector quantum efficiency.
Therefore, this definition of the true values leaves out both Poisson
fluctuations and optical attenuation in the scintillator; the latter effect is
consistent with the sign of the observed bias.
Note that this bias propagates to the proton energy \Ep calculated from this
number of photons, and ultimately therefore to the neutron energy \En and the
scattering angle $\theta$.
To address the bias, we plan to recast the likelihood to depend on
the \emph{emitted} rather than the \emph{detected} number of photons from
each interaction, which more directly relates to the deposited energy, but
requires implementing a non-trivial efficiency integral.

\begin{cfigure1c}
\includegraphics[width=0.24\textwidth]{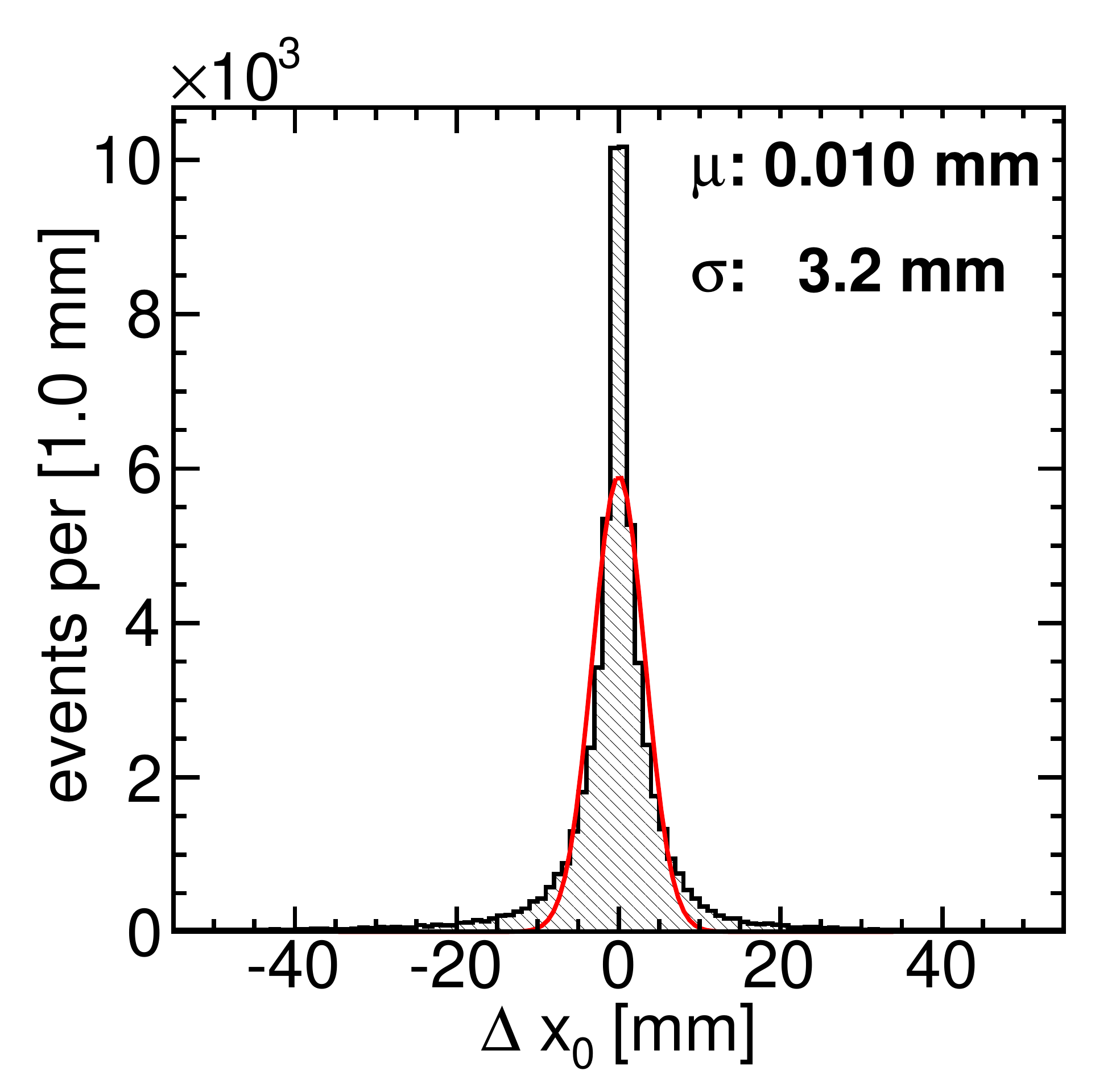}
\includegraphics[width=0.24\textwidth]{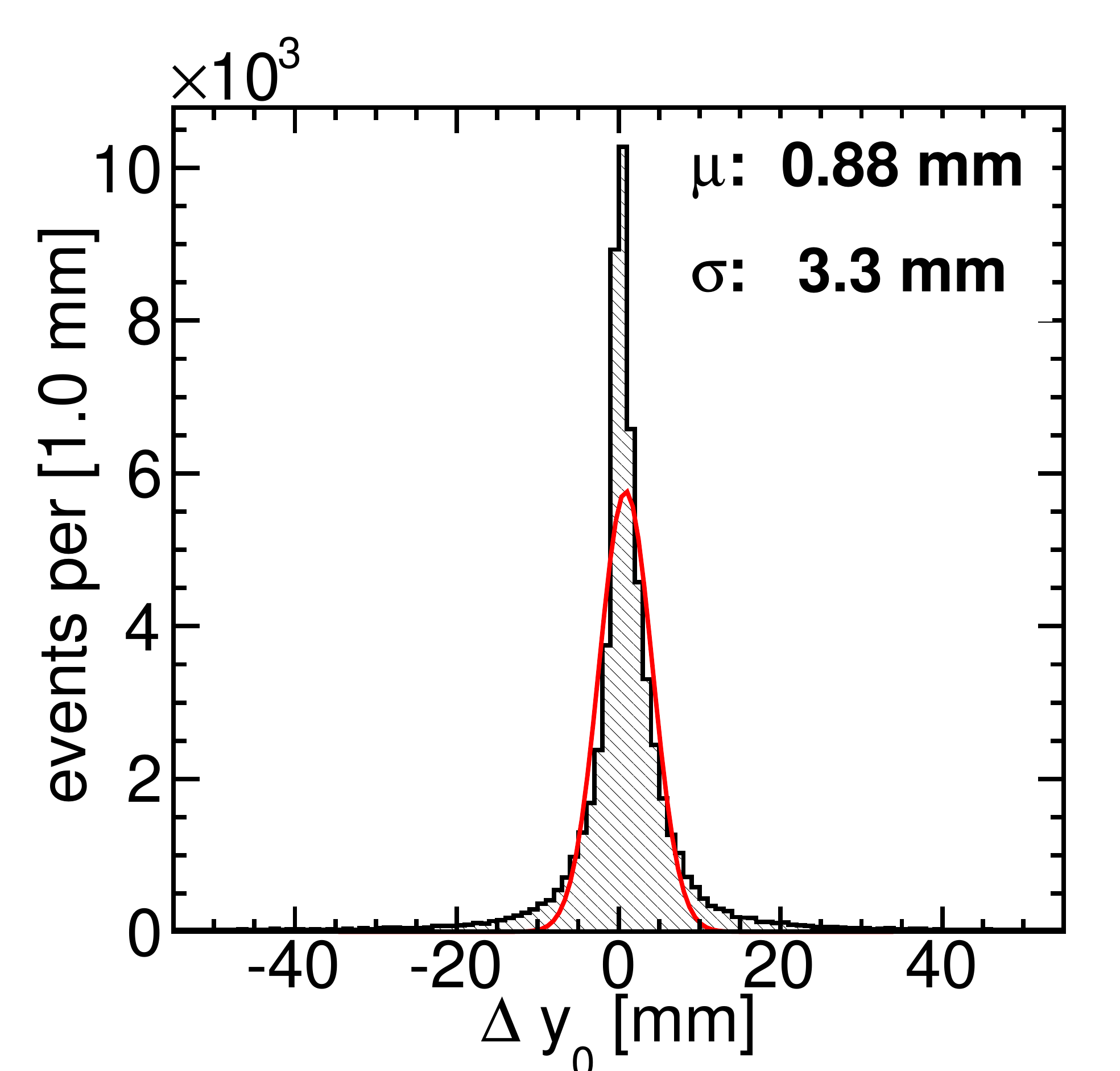}
\includegraphics[width=0.24\textwidth]{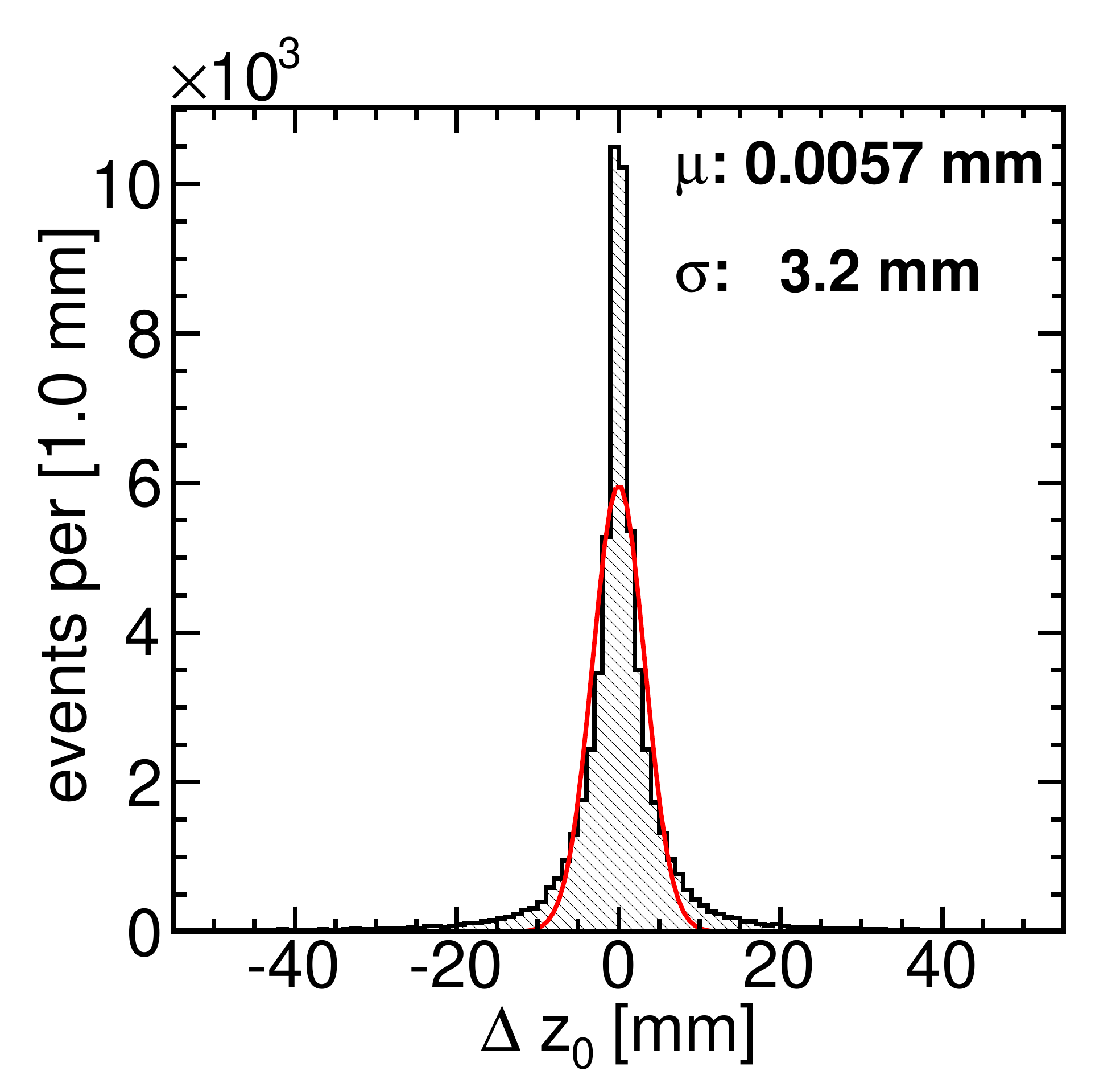}
\includegraphics[width=0.24\textwidth]{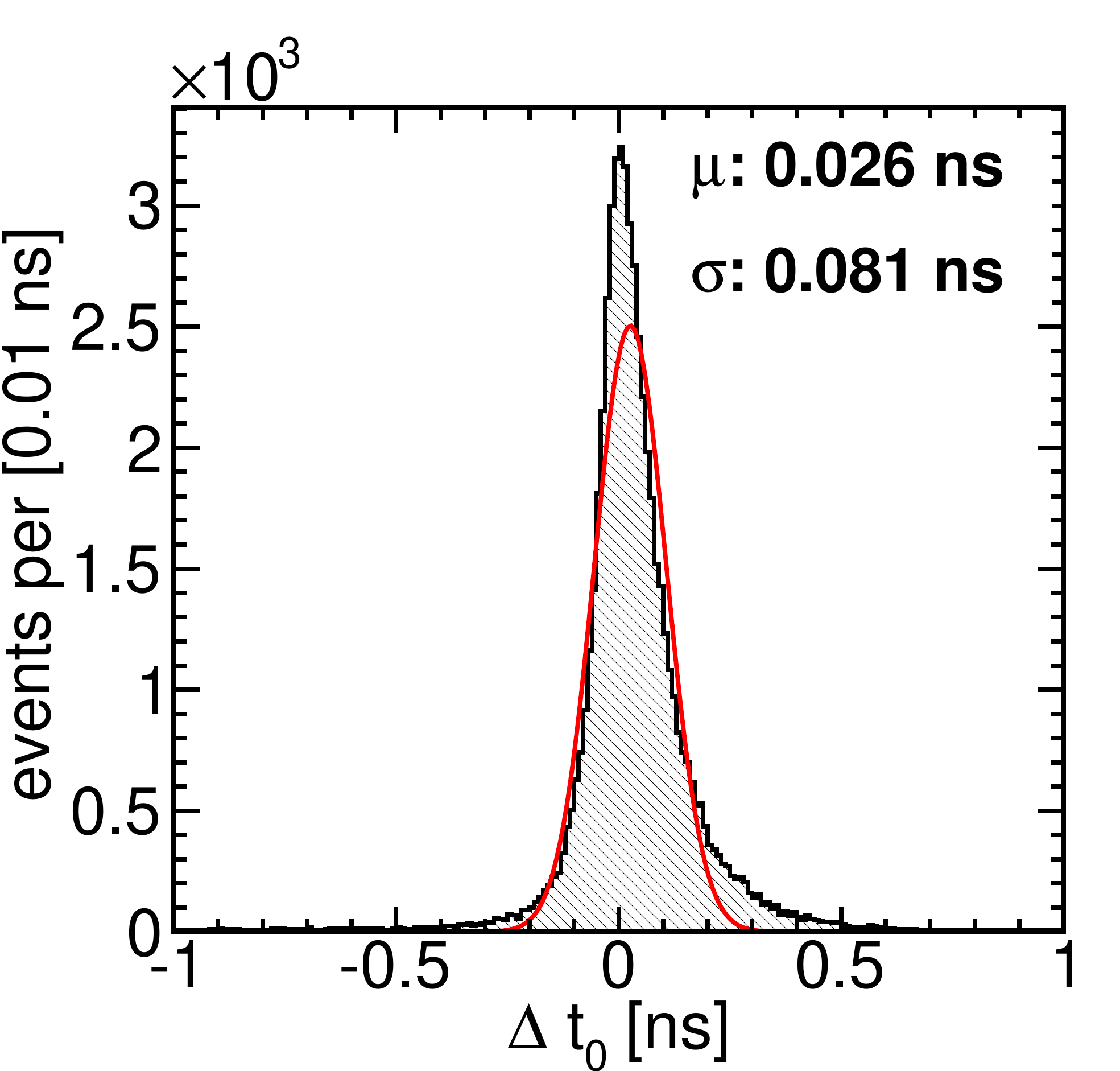}
\includegraphics[width=0.24\textwidth]{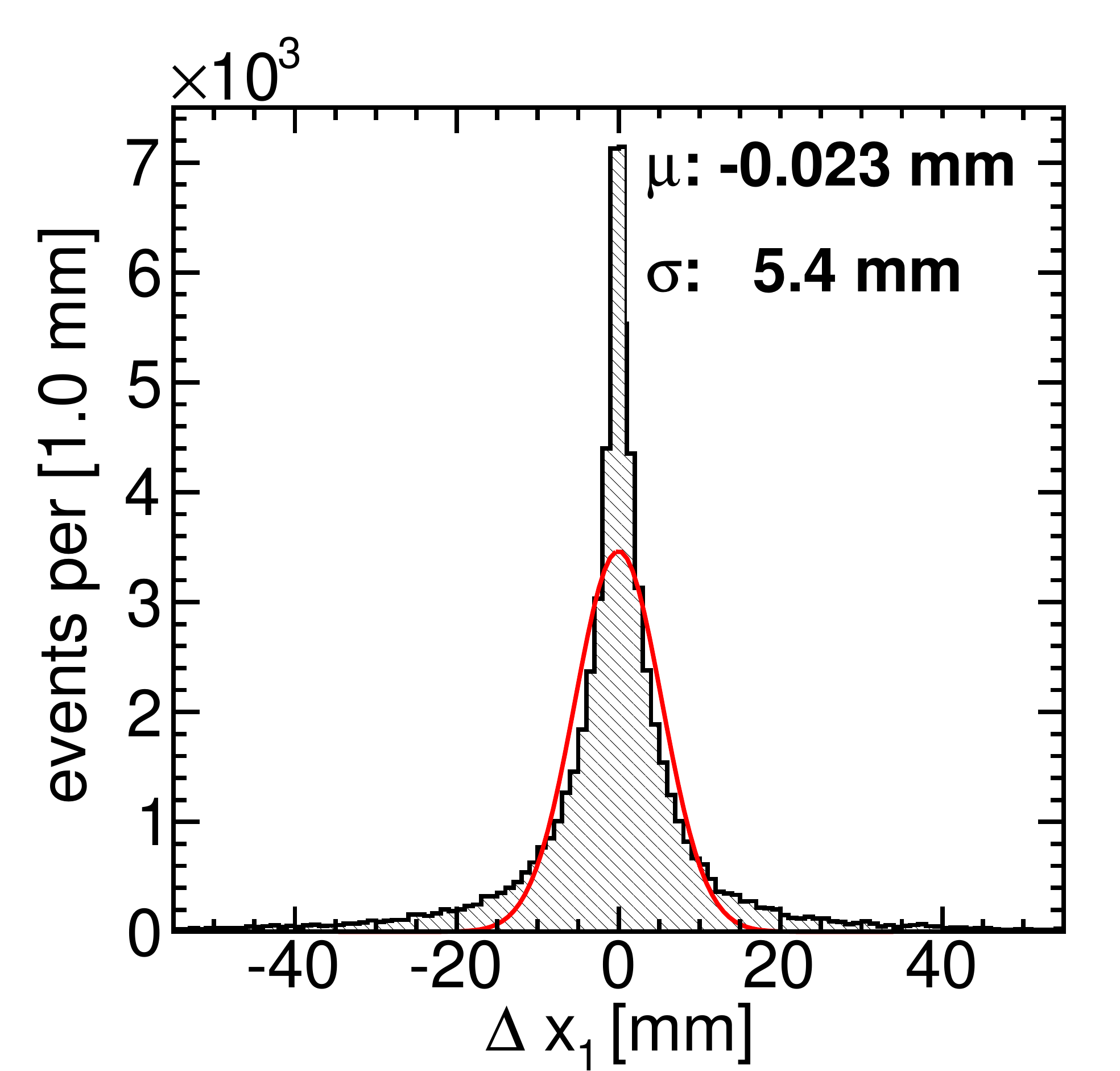}
\includegraphics[width=0.24\textwidth]{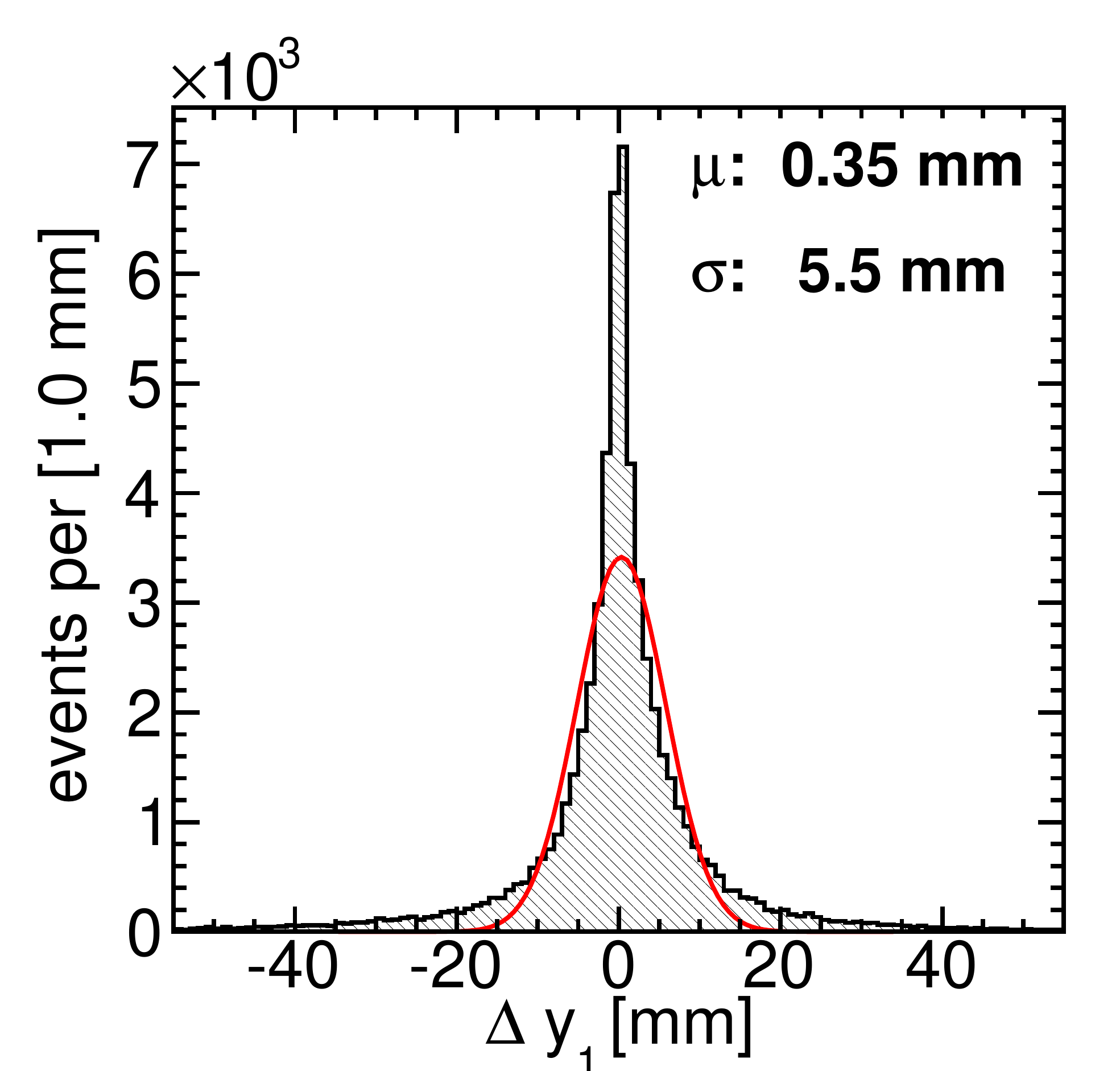}
\includegraphics[width=0.24\textwidth]{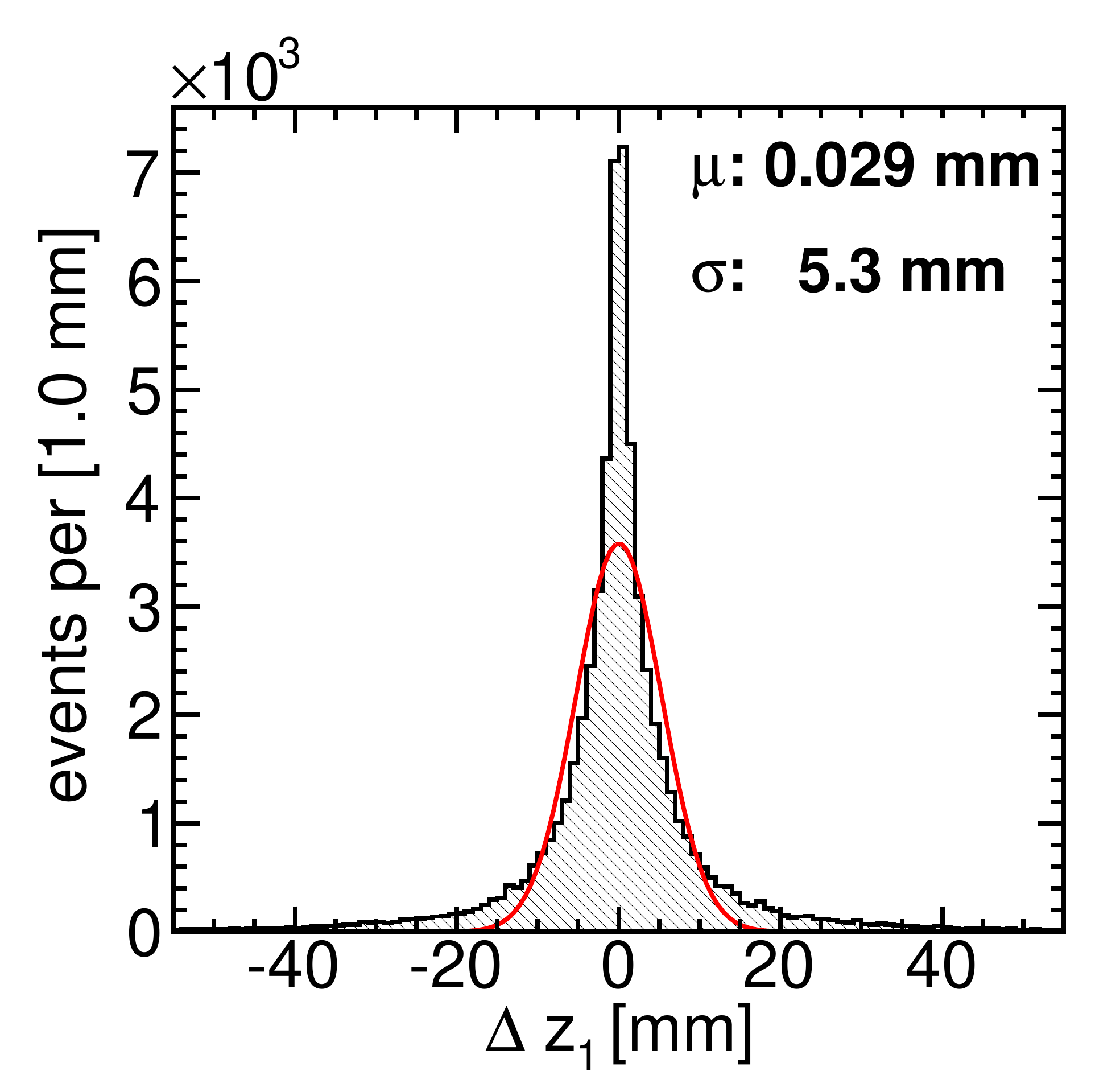}
\includegraphics[width=0.24\textwidth]{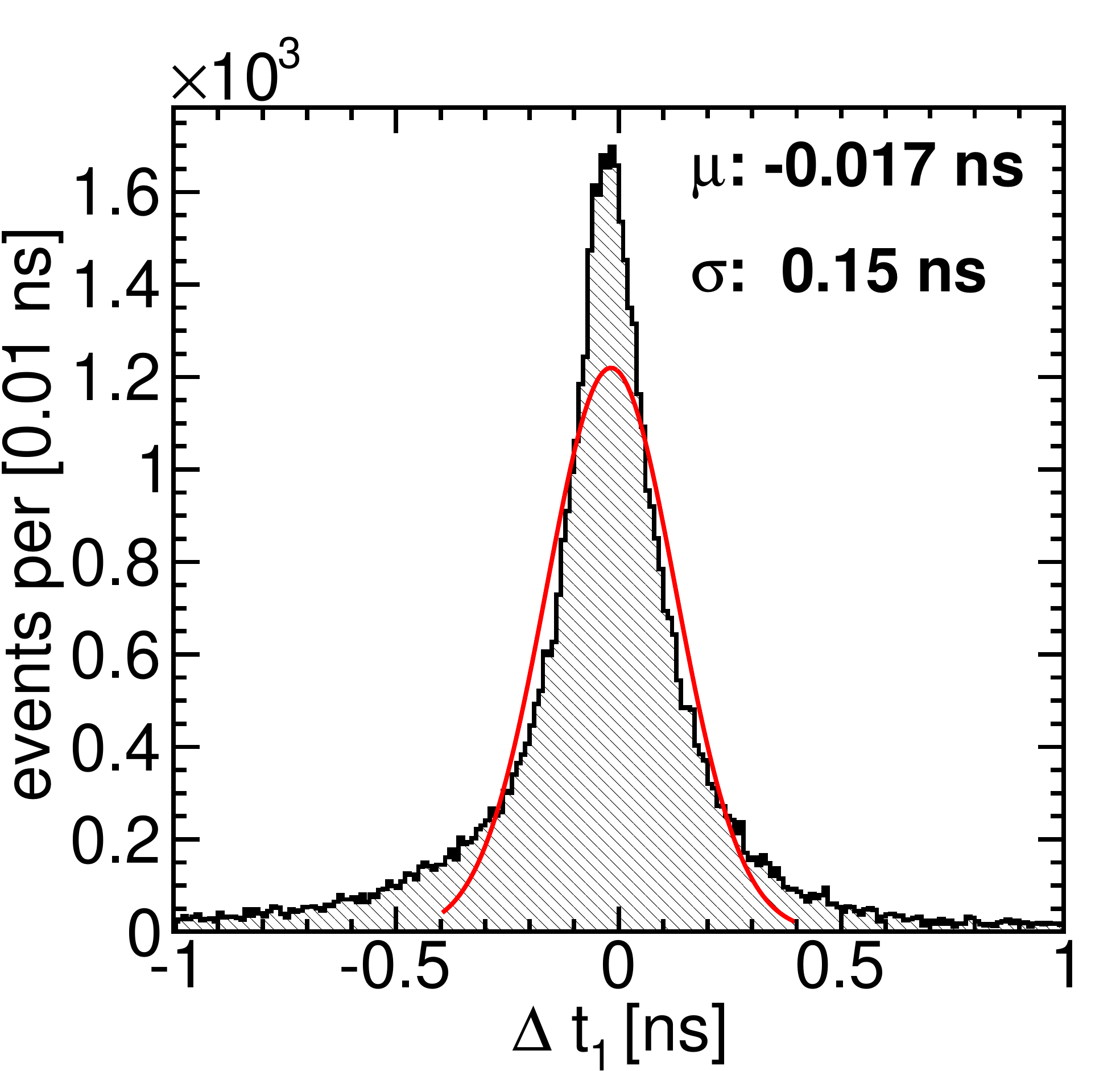}
\caption{Histograms of \Dxzero, \Dyzero, \Dzzero, \Dtzero, representing the 
accuracy of the primary reconstructed quantities for the first neutron 
interaction.}
\label{f:reco1}
\end{cfigure1c}

\begin{cfigure}
\includegraphics[width=0.49\columnwidth]{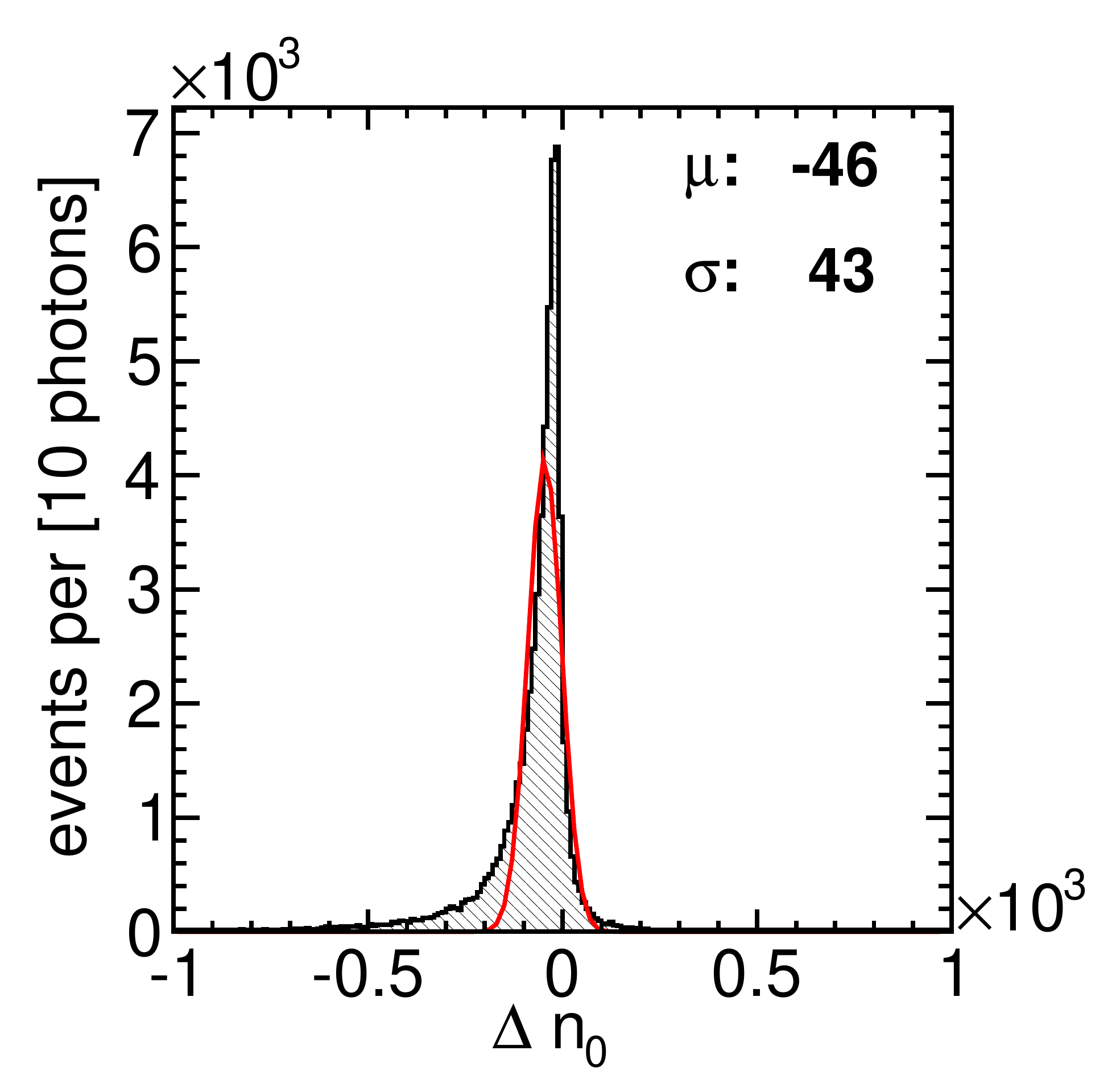}
\includegraphics[width=0.49\columnwidth]{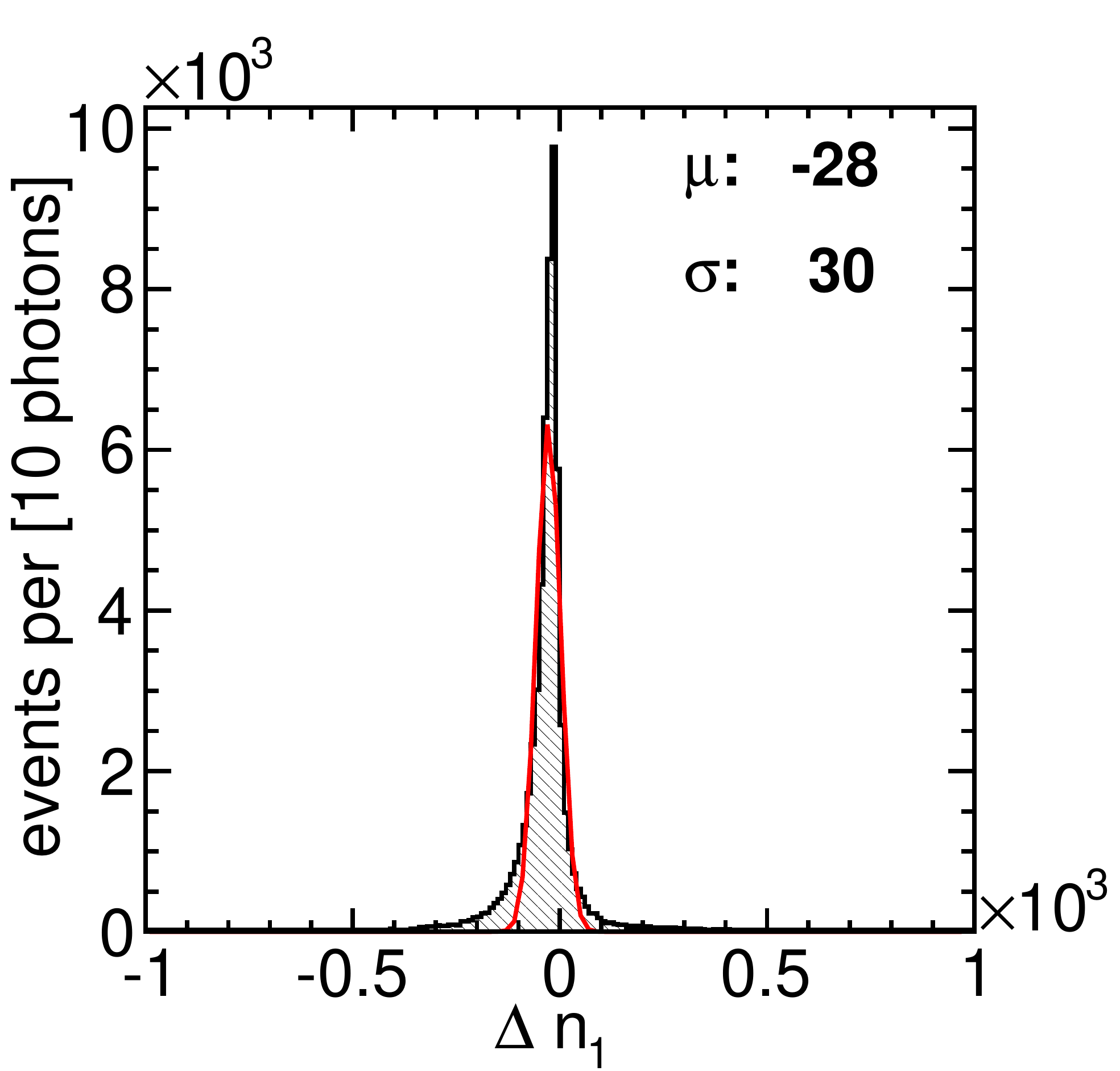}
\caption{Histograms of \Dnzero, \Dnone, representing the number of photons 
assigned to each interaction by the reconstruction, from which the deposited 
energy is calculated.}
\label{f:reco3}
\end{cfigure}

The $\Delta$-histograms of the distance \dsep and time \tsep between the first
and the second interaction, plotted in \fig{f:reco4}, are more relevant
quantities for the event reconstruction, as the individual positions and times
do not appear in the kinematic equations. The cores of these distributions are
thought to represent well reconstructed events and show Gaussian standard
deviation values of \genunit{7.1}{mm} and \genunit{170}{ps}. These quantities
are derived respectively from combining the coordinates and times of the two 
interactions, which are not independent variables but instead are correlated 
by the reconstruction algorithm. Therefore, the asymmetric non-Gaussian tails 
of the \Ddsep and \Dtsep histograms reveal correlations in the misreconstruction
of the individual interaction variables, and their impact on the reconstruction
algorithm limits is discussed below. Also included in \fig{f:reco4} is the
$\Delta$-histogram of the energy deposited by the first recoiling proton, \Ep,
calculated from $n_0$. The quantities \Ep, \dsep, and \tsep constitute the
inputs used to calculate the incoming neutron energy \En and scattering angle
$\theta$ according to \eqnrange{eq:kin1}{eq:kin3}.

\begin{cfigure1c}
\includegraphics[width=0.32\textwidth]{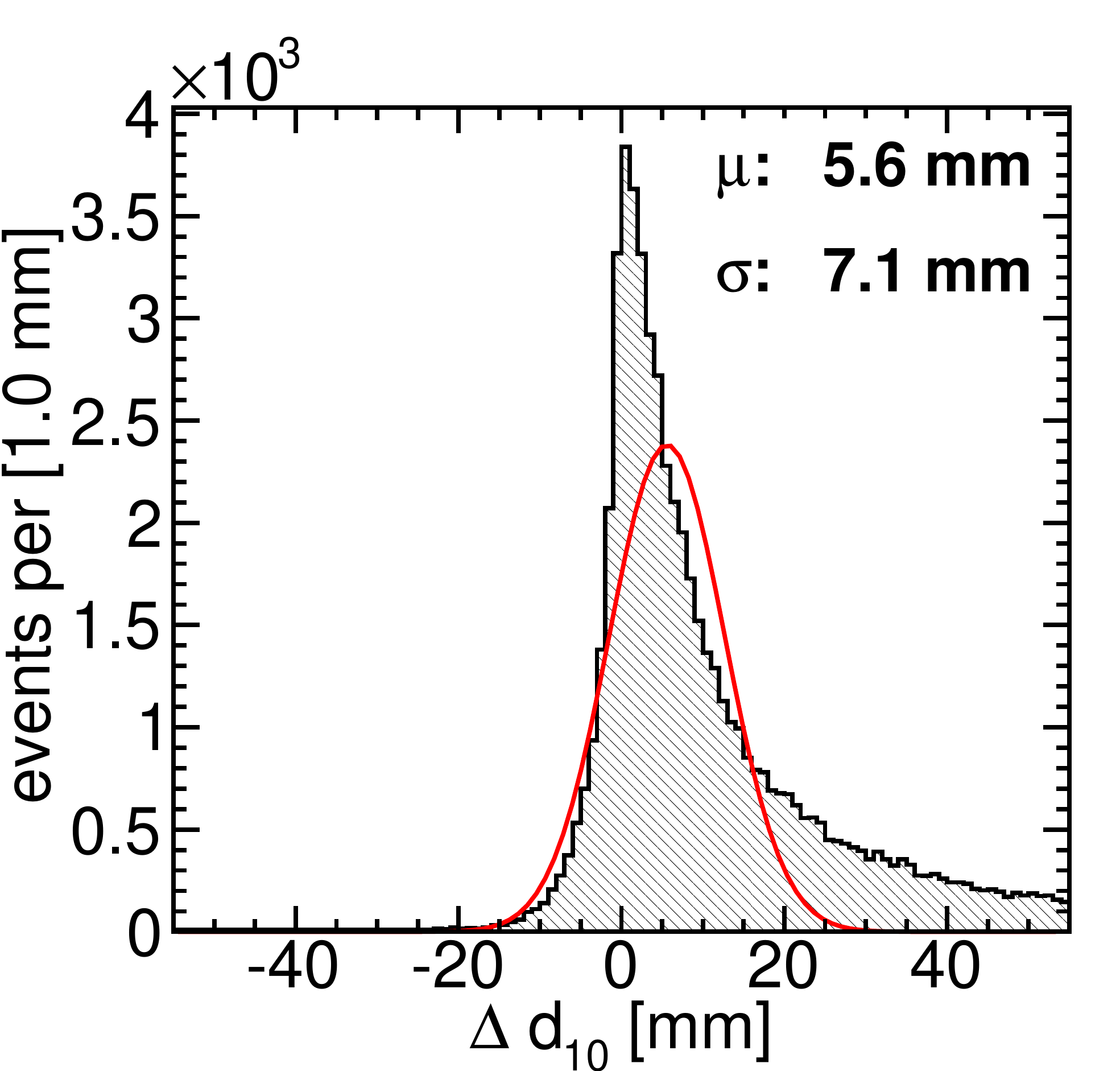}
\includegraphics[width=0.32\textwidth]{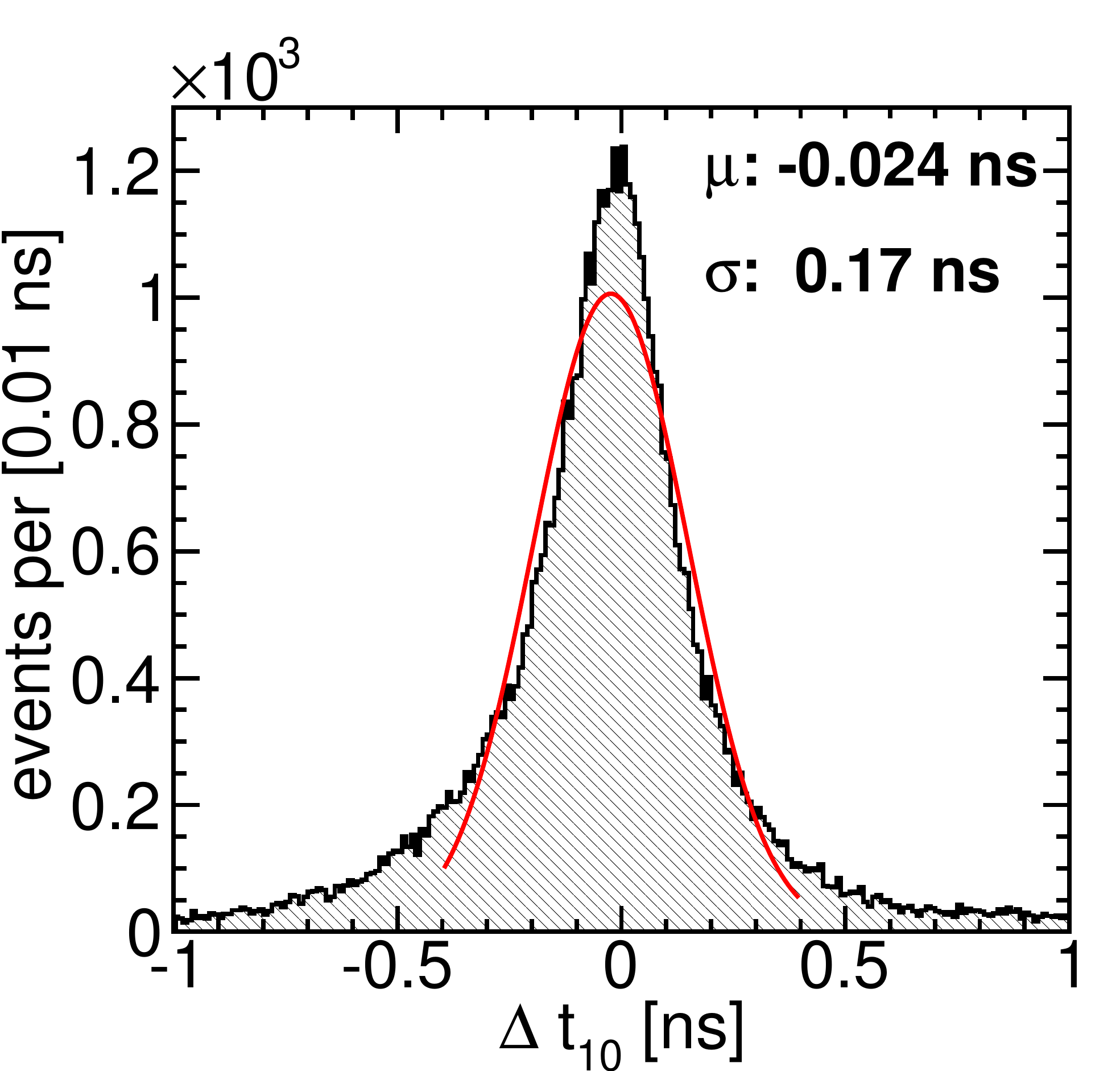}
\includegraphics[width=0.32\textwidth]{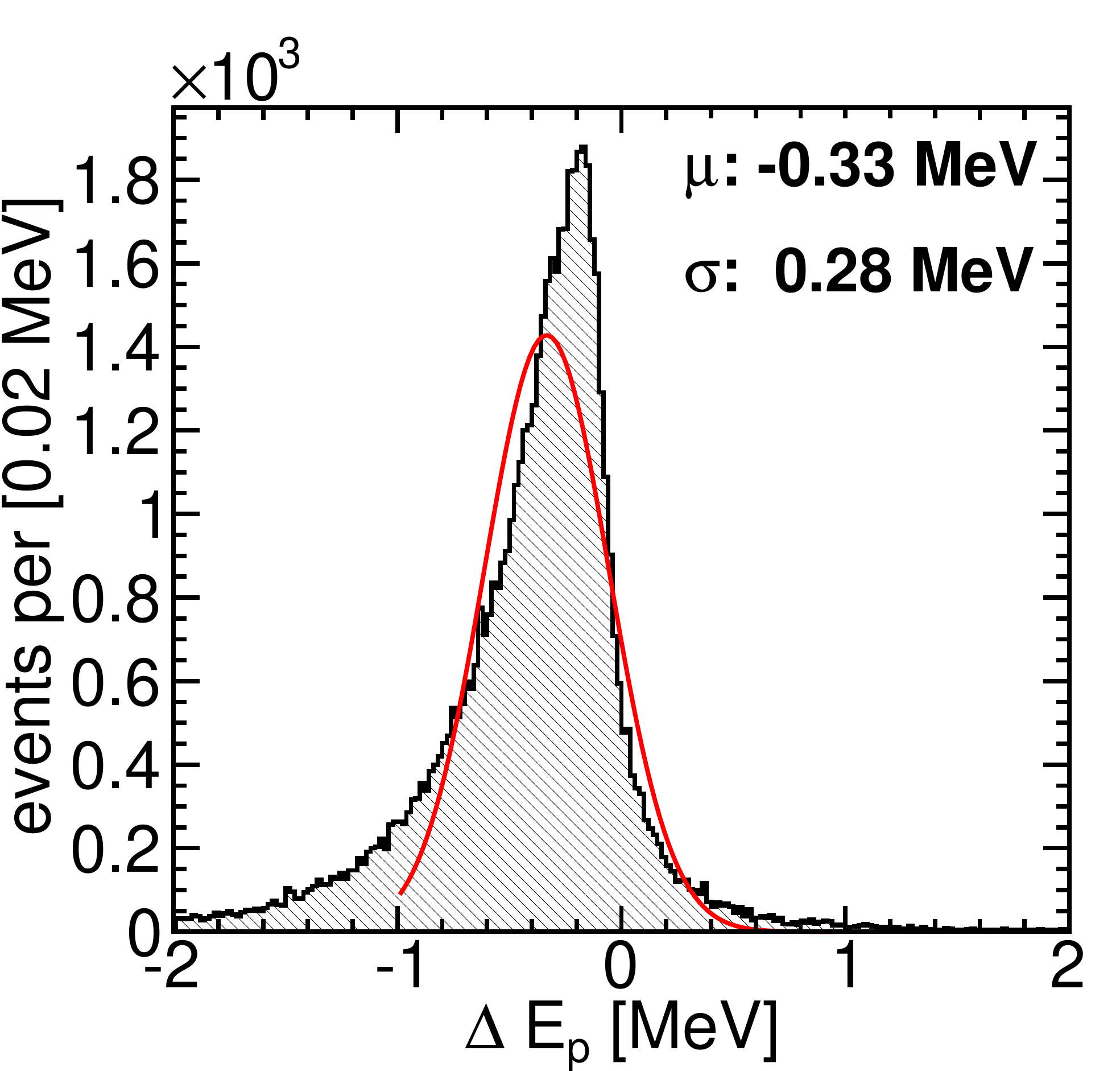}
\caption{Histograms of the intermediate derived quantities \Ddsep, \Dtsep, 
\DEp.}
\label{f:reco4}
\end{cfigure1c}

In order to evaluate the effect of the event reconstruction in the final 
reconstruction of the source spectrum and image, the histograms of \DEn and 
\Dtheta are plotted in \fig{f:comp1}. We also plot the histogram of the 
angular difference \Dalpha between the reconstructed and the true vector 
defining the axis of the incoming neutron cone, since errors in the cone 
axis direction will also effect the image reconstruction along with 
errors in the cone opening angle $\theta$. Note that \Dalpha is unique among
these resolution plots in being an \emph{unsigned} angular difference, and
consequently we do not expect a symmetric distribution peaked at zero; therefore
we do not include a Gaussian fit for \Dalpha. As for a Rayleigh distribution,
we use the mode of the distribution as an estimate of the underlying angular
uncertainty in each of two independent orthogonal components; this is the value
given in \tab{t:perf}. However, it should be noted that the \Dalpha distribution
has a much longer tail than the corresponding Rayleigh distribution, so as in
the case of the Gaussian widths, this value at best represents the ``core'' of
well reconstructed events.

\begin{cfigure1c}
\includegraphics[width=0.32\textwidth]{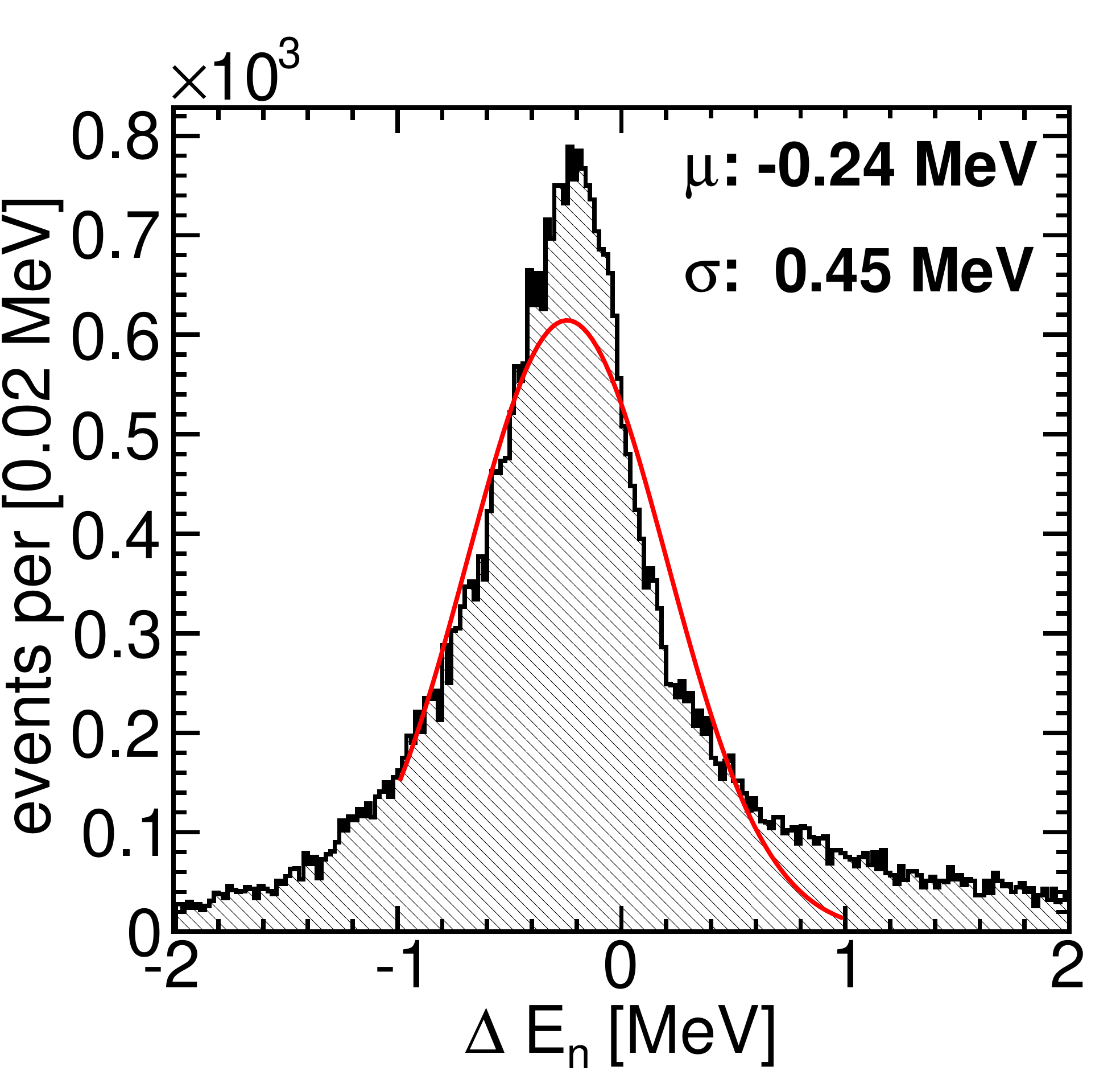}
\includegraphics[width=0.32\textwidth]{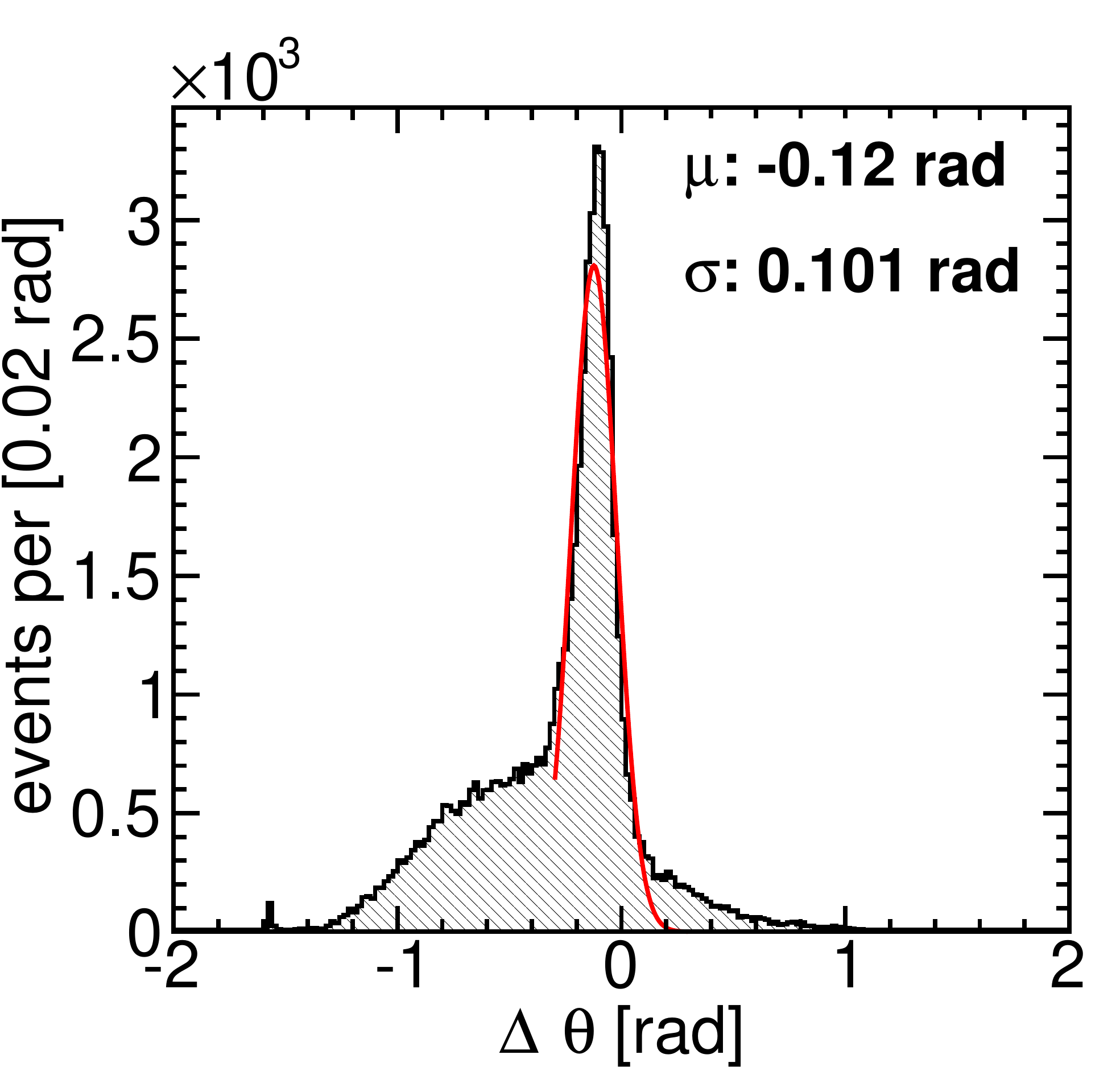}
\includegraphics[width=0.32\textwidth]{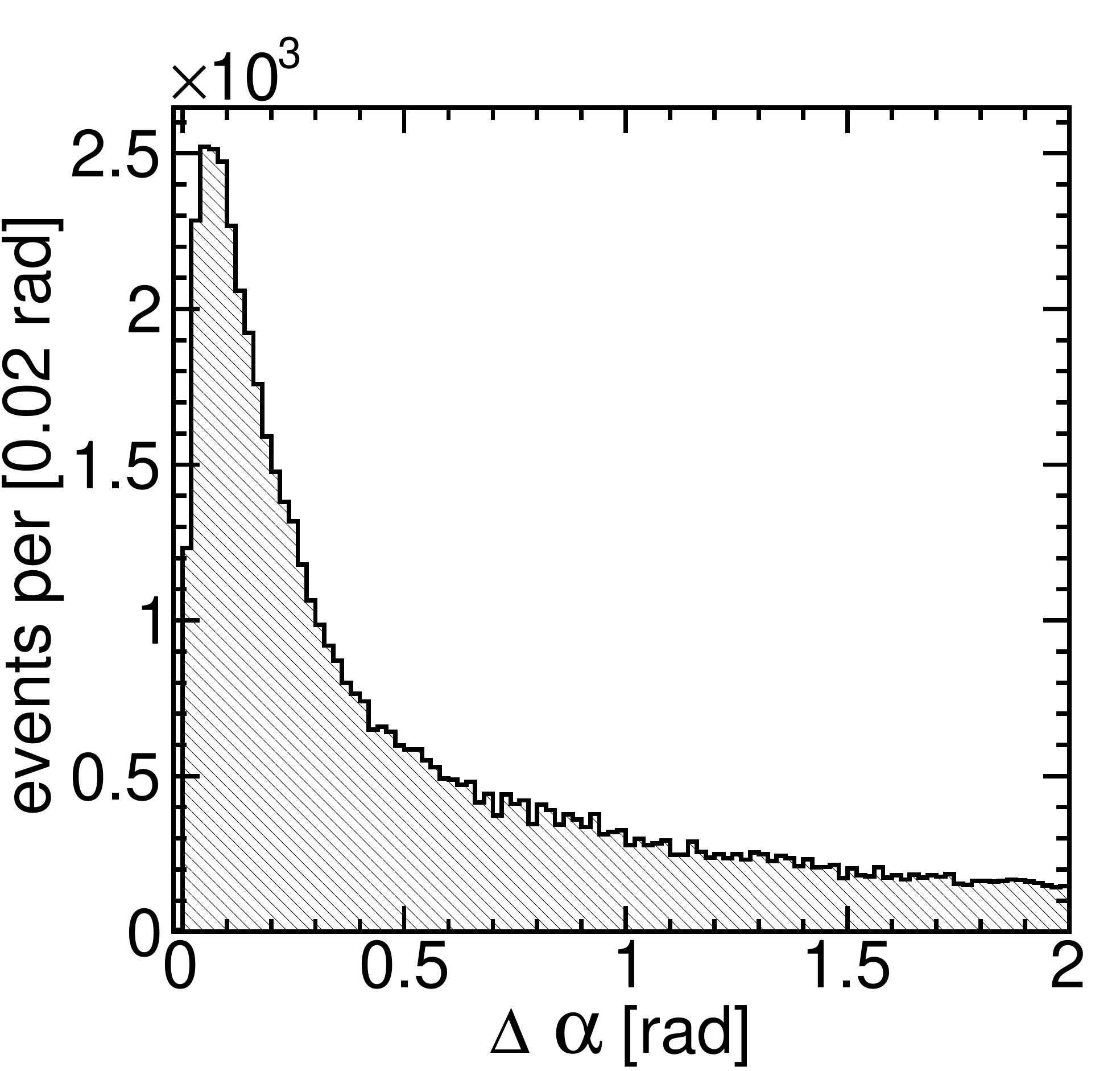}
\caption{Histograms of the final derived quantities used for spectrum and 
image reconstruction: \DEn, \Dtheta, \Dalpha.}
\label{f:comp1}
\end{cfigure1c}

The most salient feature of 
these histograms is the broad peak for negative values of \Dtheta 
representing reconstructed scattering angles smaller than the true 
angles.
These misreconstructed events correlate with the extended high tail of \Dalpha
values,
as well as other tails in the previous plots of reconstructed quantities.
Further investigation indicated that these misreconstructed events are primarily
observed when the distance between the first two neutron interactions in the
scintillator is small, i.e.\ less than about \cm{1.5}.
\Fig{f:comp3} shows \Dtheta as well as \Ddsep plotted against the true \dsep
value, illustrating that result.
These events also of course tend to have small time difference between the first
two interactions.
One likely explanation for this observation, as mentioned above, is that when
the first two interactions are close in space and time, they are reconstructed
as a single interaction; the reconstruction algorithm then identifies a
subsequent interaction or group of interactions as the second one.

\begin{cfigure}
\includegraphics[width=0.99\columnwidth,height=0.4\textheight,keepaspectratio]{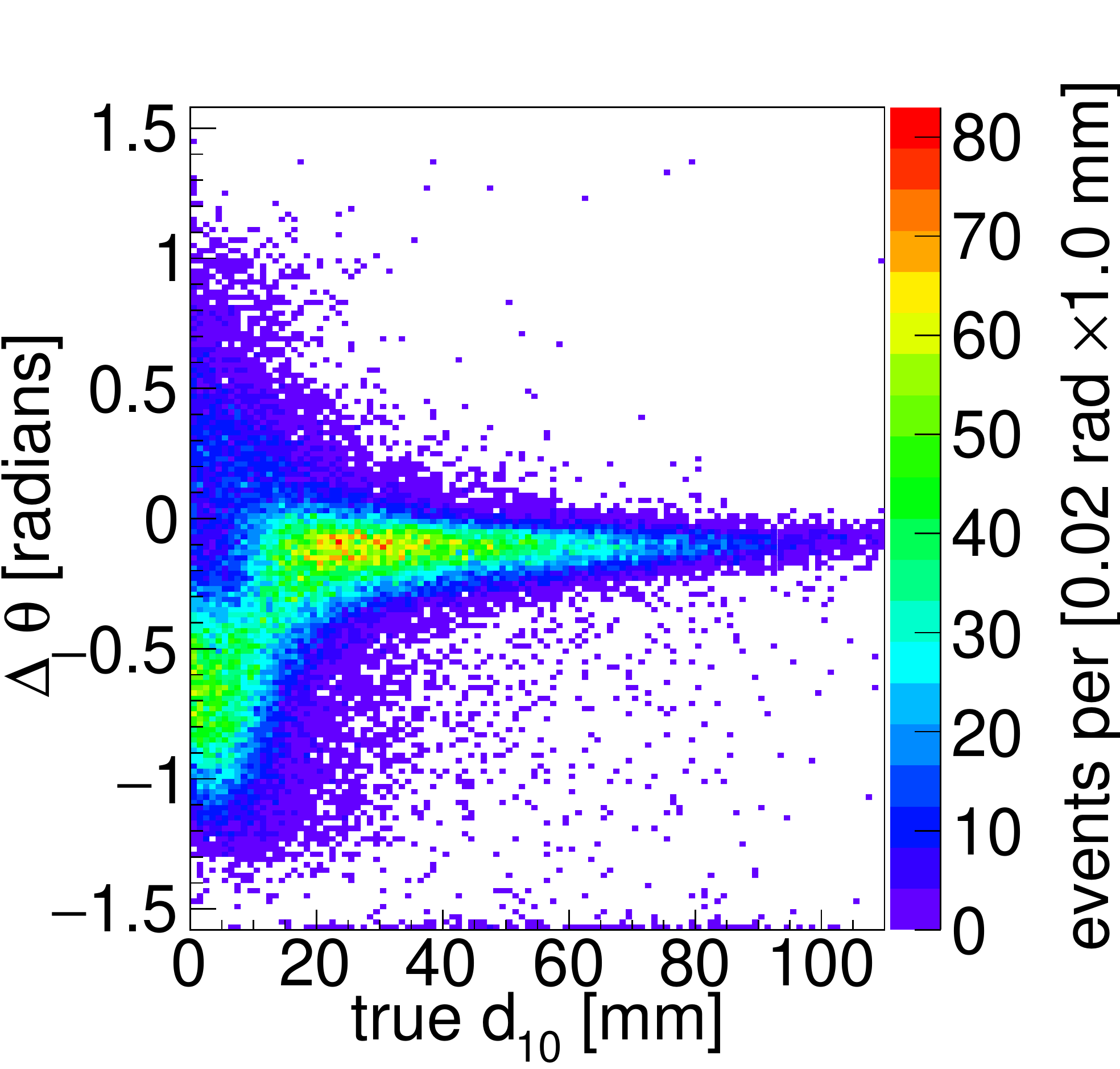}
\includegraphics[width=0.99\columnwidth,height=0.4\textheight,keepaspectratio]{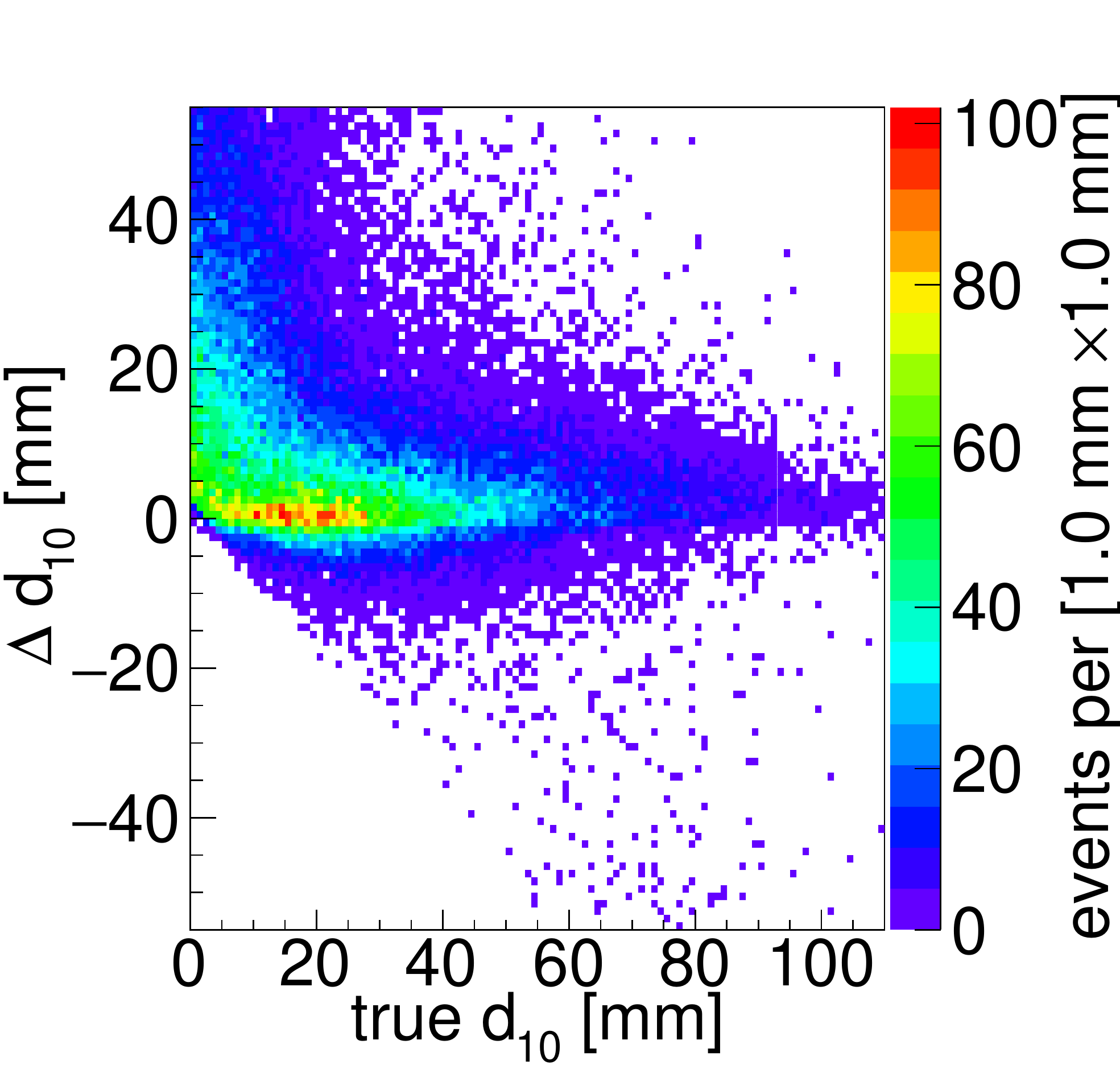}
\caption{
Histograms of \Dtheta vs true \dsep, and \Ddsep vs true \dsep. These plots
demonstrate that most of the misreconstructed events discussed in the text
occur when the first two neutron interactions are close together ($<\cm{1.5}$).}
\label{f:comp3}
\end{cfigure}

\Fig{f:smear} provides a graphical comparison of the effect that smearing the 
photon time ($T$) and position coordinates $(X,Y,Z)$ have in the reconstruction quality, 
also showing the intermediate cases in
which only time or only position resolutions were applied.
The figure demonstrates that the time resolution is the main contributor to the
error in event reconstruction; it also reduces the reconstruction success rate
by 20\%--30\%.
In contrast, the smearing in 
photon detection position within the pixel size of currently available MCP-
PMT technology has little impact on the event reconstruction.
However, the pixel size drives not only position uncertainty but also photon
occupancy per pixel, which does not affect these
results but will be an important consideration for a hardware implementation.

\begin{cfigure}
\includegraphics[width=0.99\columnwidth]{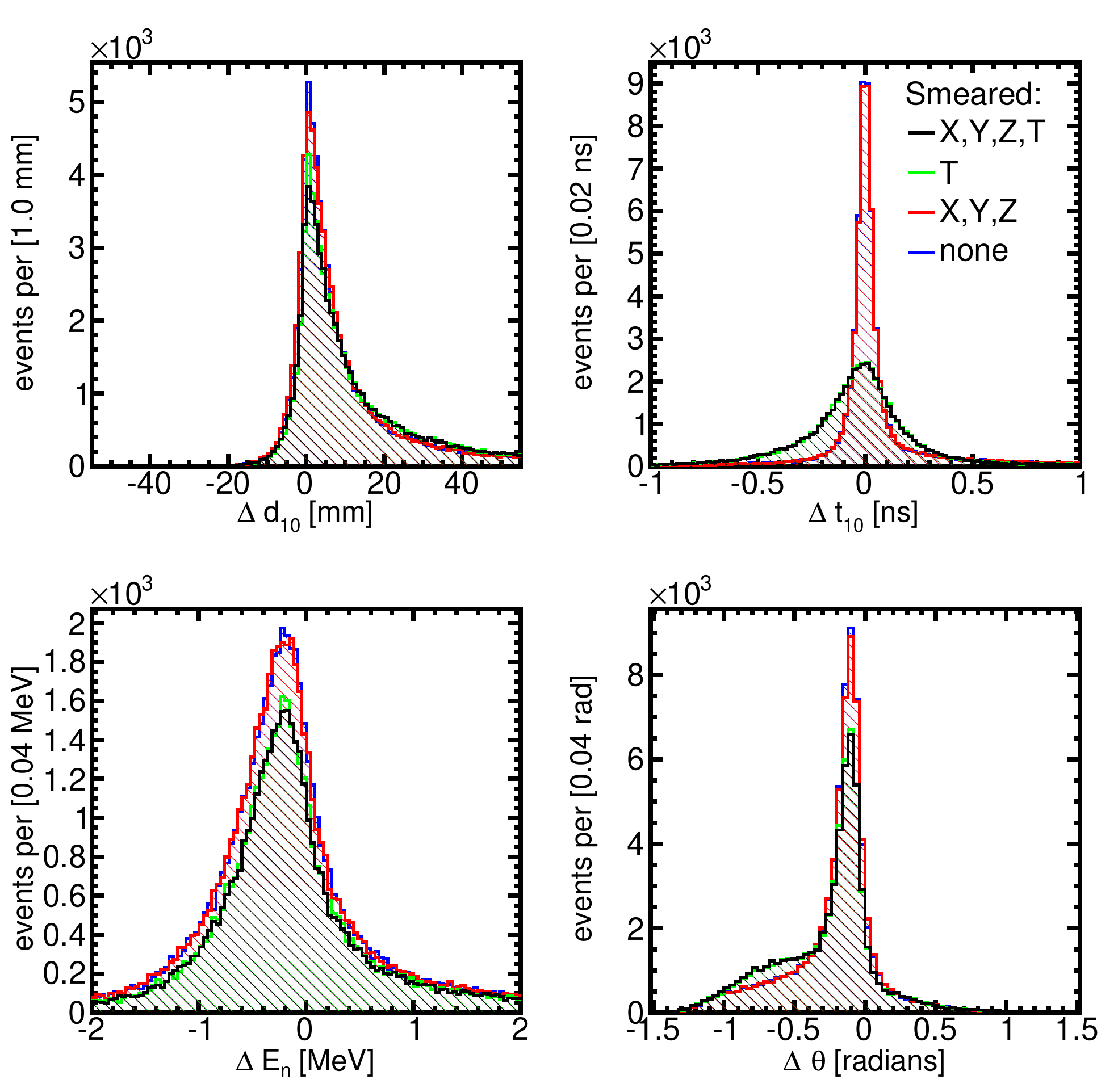}
\caption{Histograms comparing the reconstruction performance with photodetector
spatial and temporal resolution included and excluded from the reconstruction
inputs. Upper left: \Ddsep. Upper right: \Dtsep. Lower left: \DEn.
Lower right: \Dtheta.}
\label{f:smear}
\end{cfigure}

Furthermore, 
when exact photon times are used, the negative tail in the \Dtheta histogram 
decreases in size,  and events with the first two interactions as close as
\cm{\sim 1} and \genunit{\sim 0.6}{ns} can be successfully reconstructed.
While this indicates that reduction in the photon 
detection time resolution beyond \genunit{0.1}{ns} can potentially improve the
SVSC performance further, it also establishes an intrinsic limit 
in event reconstruction resolution within this algorithm framework.

Finally, we consider the highest-level reconstructed distributions, the energy
spectrum and image. A reconstructed source energy spectrum is presented in 
\fig{f:recoSpectrum}, where it is compared with the ``true'' spectrum 
calculated using the known true values of the interactions' location, time and 
deposited energy, as well as with the actual incident energy spectrum of the
same set of events. The latter two distributions are quite similar albeit with
some differences at low \En, but the reconstructed energy spectrum has two
qualitative issues. First, for the accurately reconstructed events in the core
of the spectrum, \En is biased low due to the absence of a correction for optical
attenuation as discussed above. Second, there is a long tail of misreconstructed
events with high \En; in fact, fully 20\% of the reconstructed events overflow
this plot ($\En>\mev{20}$).

\begin{cfigure}
\includegraphics[width=0.99\columnwidth]{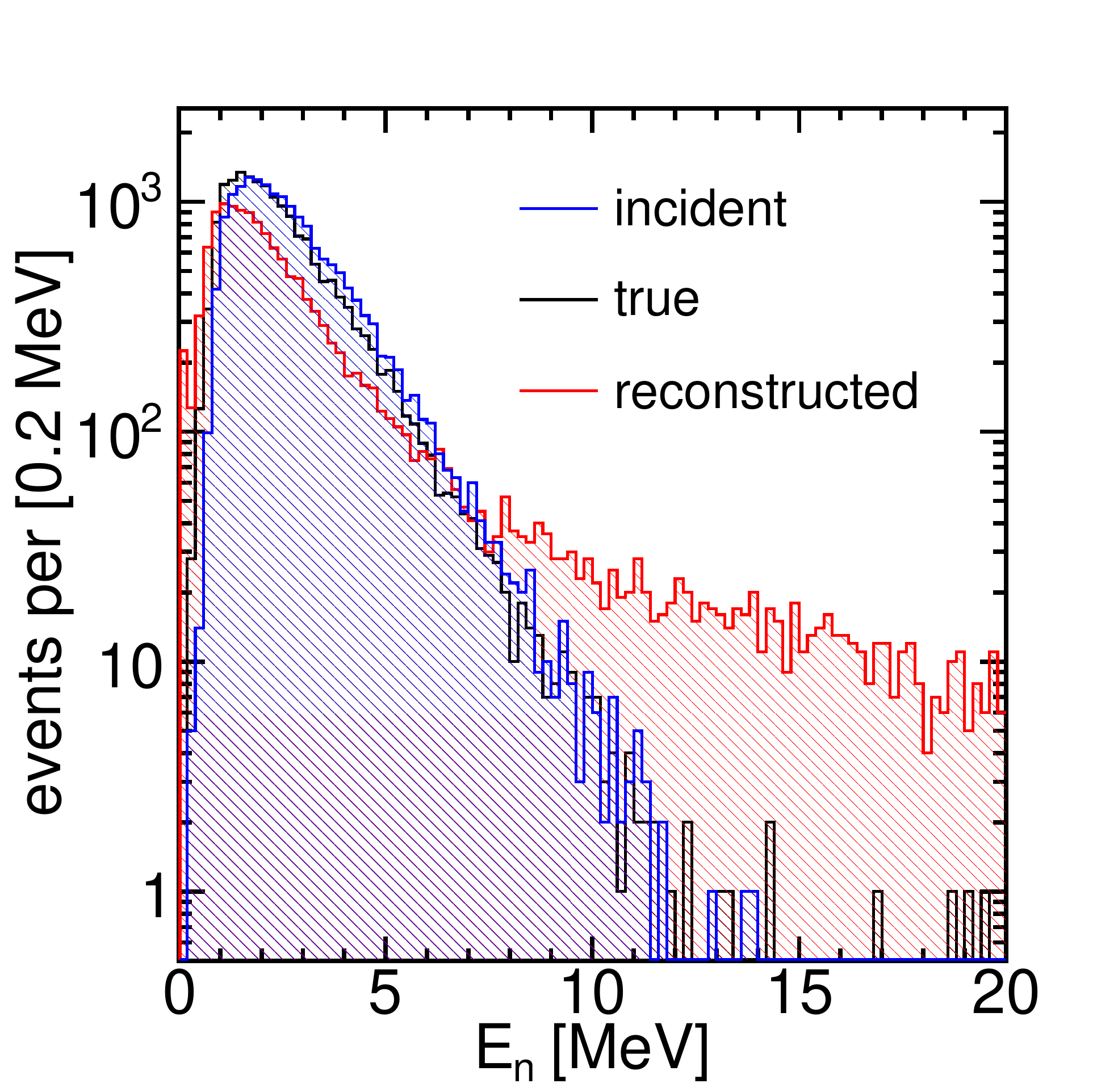}
\caption{Reconstructed neutron energy spectrum for the simulated \Cftft source.
The red spectrum is derived from the reconstructed quantities, while the black
spectrum is, for the same events, the spectrum that would be calculated from
the true positions and times of the first two proton recoils above
threshold and the true energy deposited by the first. Both spectra are sculpted 
by event selection requirements, and by
errant event topology such as a carbon scatter between the first two hydrogen
scatters. Also included is the spectrum of the actual energy 
of the incident neutrons of the same events entering the previous two spectra, which 
is sculpted by the same event selection but not topological effects.}
\label{f:recoSpectrum}
\end{cfigure}

The reconstructed image of \fig{f:recoImage} shows good agreement with the 
simulated source location, indicated by a black cross in the lower zoomed-in
view. In the case of the image, misreconstructed events tend to contribute
randomly, which reduces their importance in determining a source centroid.
This image was reconstructed with list-mode MLEM under the assumption of a
simple system response, in which for example the two reconstructed interactions
are always taken to be the first two neutron interactions. This method produces
qualitatively better results than simple backprojection. However, we expect that
further improvements could be obtained in the future with a more complete system
response function, including better treatment of experimental resolutions and 
alternate event topologies, such as, for example, including the possility of an 
initial carbon scatter. Quantification of the 
SVSC image quality for point and extended sources will be done in future work.

\begin{cfigure}
\includegraphics[width=0.99\columnwidth,height=0.4\textheight,keepaspectratio]{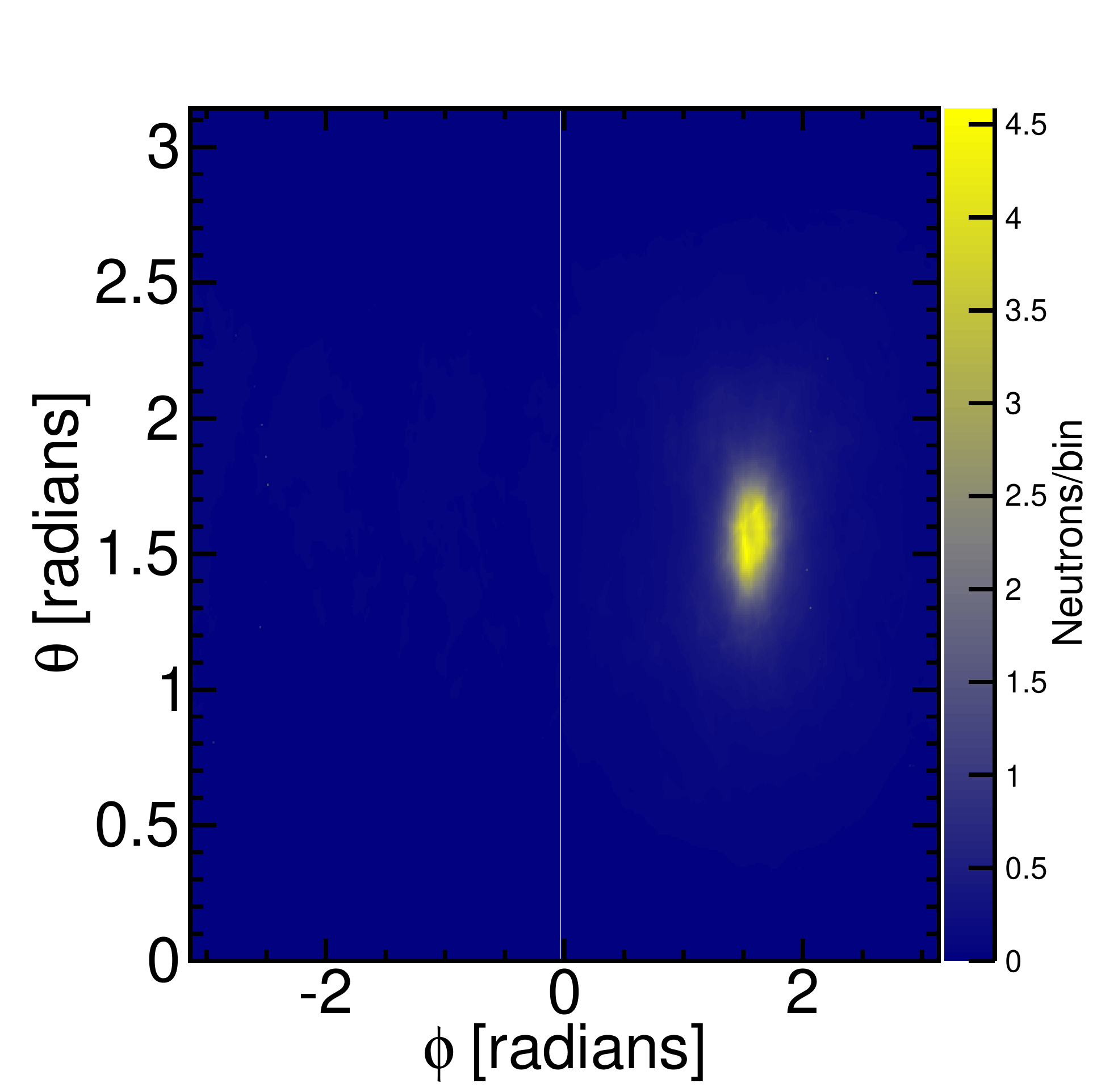}
\includegraphics[width=0.99\columnwidth,height=0.4\textheight,keepaspectratio]{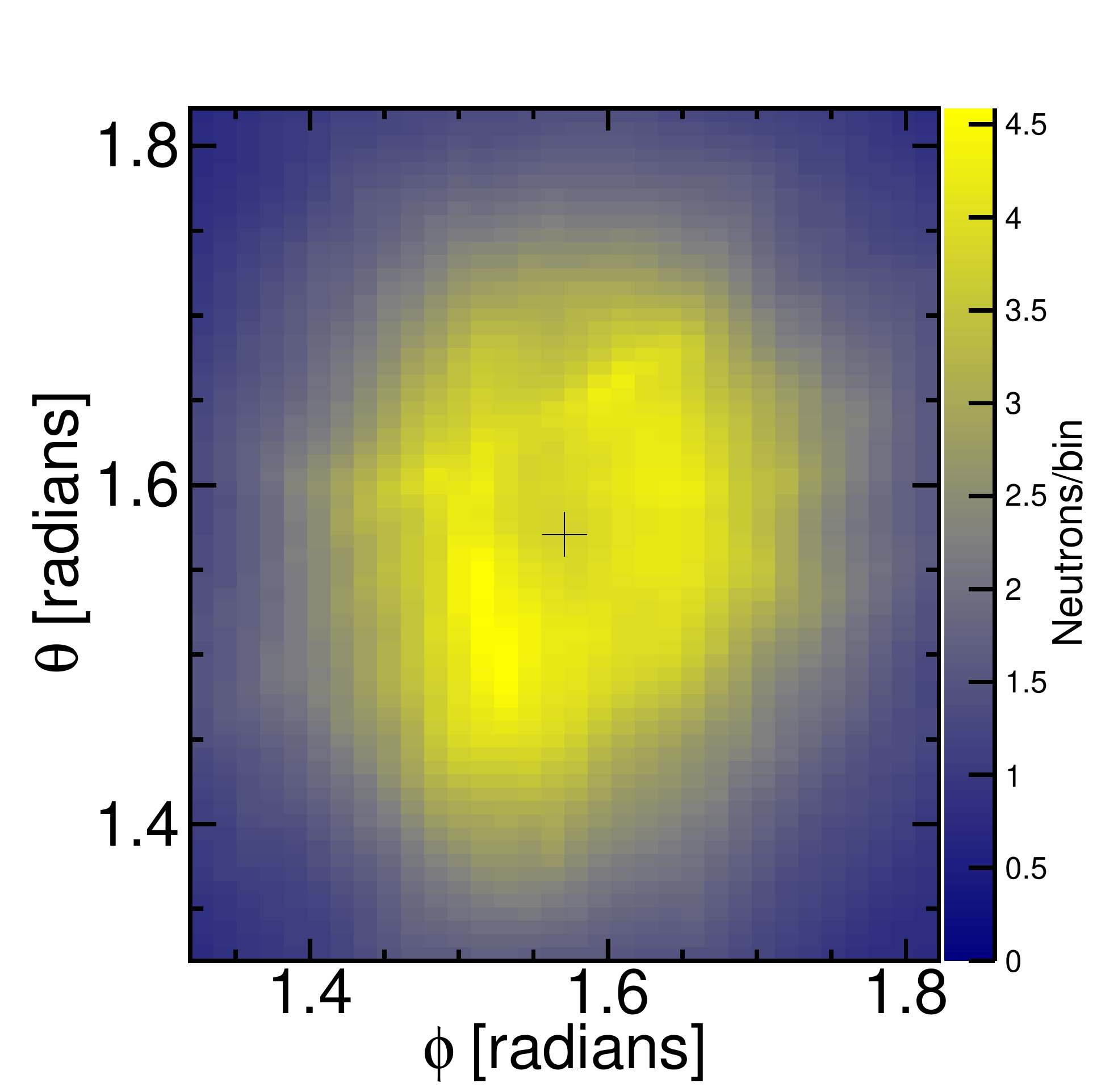}
\caption{Reconstructed image for the simulated \Cftft source. Top: image 
reconstructed over the full $4\pi$ space. Bottom: zoomed in to the centroid 
region.  The black cross represents the simulated source location. The color
scale indicates the number of imaged neutrons from each bin in this source
space; there are $400\times400$ bins in the two-dimensional image.}
\label{f:recoImage}
\end{cfigure}

\section{Conclusions}
\label{s:conclusions}
We have described the theoretical advantages of single-volume double-scatter 
neutron imaging, namely a high imaging efficiency and a compact form factor. 
We also demonstrated the feasibility of the direct reconstruction SVSC 
concept via Monte Carlo simulation and reconstruction of a neutron point 
source.

For the simulated detector system, the results indicate about 13.6\% of incident
fission-energy neutrons result in at least two proton scattering interactions
above \kev{300}, with about half of those events successfully
reconstructed using the described event likelihood maximization algorithm.
A core resolution of \genunit{3}{mm} (\genunit{5}{mm}) in each spatial dimension
and \genunit{80}{ps} (\genunit{150}{ps}) in time
is observed for the first (second) interaction, while a
limitation of the current likelihood expression results in a bias in the
reconstructed number of photons for each interaction. Resolutions of
higher-level quantities, such as the distance between the two interactions, have
complicated distributions, indicating the presence of non-trivial correlations
among the low-level quantitites.

Most of the poorly reconstructed events have small distance (and time) between
the two first interactions; this is expected both because these events are more
difficult to reconstruct and because, even for a given spatial/temporal
resolution, they contain less neutron energy and direction information due to
the reduced lever arm.

The response of the system to gammas has not been addressed here, but is an
important part of ongoing simulation studies. In particular we would like to
understand the ability of the
system to discriminate neutron from gamma events based on interaction timing
in the absence of pulse shape discrimination.

Physical construction, calibration, and operation of the simulated system 
presents several technical challenges, and the next desired step is 
demonstration of the SVSC concept in a laboratory prototype system.

\section*{Acknowledgements}
We thank Glenn Jocher and Kurtis Nishimura for sharing knowledge and wisdom 
from their experience with a closely 
related detector system. We thank Marc Ruch for his software 
implementation of list-mode MLEM for double-scatter imaging, as well as for 
various fruitful discussions on compact neutron imaging.

Sandia National Laboratories is a multimission laboratory managed and operated by National Technology and Engineering Solutions of Sandia LLC, a wholly owned subsidiary of Honeywell International Inc., for the U.S. Department of Energy’s National Nuclear Security Administration under contract DE-NA0003525.

\section*{References}

\bibliography{../svscbibfile,./thispaperextras}

\end{document}